\documentclass{ws-ijmpa}

\usepackage{epsfig}

\newcommand{\be}{\begin{eqnarray}}
\newcommand{\ee}{\end{eqnarray}}
\newcommand{\bea}{\begin{eqnarray}}
\newcommand{\eea}{\end{eqnarray}}

\newcommand{\beq}{\begin{equation}}
\newcommand{\eeq}{\end{equation}}
\newcommand{\nn}{\nonumber}

\def\la{\mathrel{\mathpalette\fun <}}
\def\ga{\mathrel{\mathpalette\fun >}}
\def\fun#1#2{\lower3.6pt\vbox{\baselineskip0pt\lineskip.9pt
\ialign{$\mathsurround=0pt#1\hfil##\hfil$\crcr#2\crcr\sim\crcr}}}

\newcommand{\up}{^\uparrow}
\newcommand{\dw}{^\downarrow}

\def\kaz{k_{1z}}
\def\kbz{k_{2z}}

\def\kap{k_{1+}}
\def\kbp{k_{2+}}

\def\kam{k_{1-}}
\def\kbm{k_{2-}}

\begin{document}

\markboth{A.V. Anisovich et al.}
{Quark-Diquark Systematics of Baryons.}

\catchline{}{}{}{}{}
%
%%%%%%%%%%%%%%%%%%%%%%%%%%%%%%%%%%%%%%%%%%%%%%%%%%%%%%%%%%%%%%%%%%%%

\title{ QUARK--DIQUARK SYSTEMATICS OF BARYONS
AND THE $SU(6)$ SYMMETRY FOR  LIGHT  STATES }

\author{A.V. ANISOVICH,
        V.V. ANISOVICH\footnote{anisovic@thd.pnpi.spb.ru}, M.A. MATVEEV,
        V.A. NIKONOV,
        A.V. SARANTSEV and T.O. VULFS}
\address{Petersburg Nuclear Physics Institute,
Gatchina, 188300 Russia}

\maketitle

\begin{history}
\received{Day Month Year}
\revised{Day Month Year}
\end{history}

\begin{abstract}
We continue our attempts to systematize baryons, which are composed of  light quarks
($q=u,d$),
as  quark--diquark systems.
The notion of two diquarks is used:
(i) $D^{1}_{1}$, with the spin $S_D=1$ and isospin
 $I_D=1$
 and (ii)  $D^{0}_{0}$, with $S_D=0$ and
 $I_D=0$. Here we try to resolve the problem of the low-lying
$\Delta(\frac 52^-)$ states: in the last experiments the lightest
state is observed at $\ga 2200$ MeV, not at $1900 - 2000$ MeV as it has been stated 20
years ago.  We are looking for different  systematization variants with the forbidden
low-lying $\Delta(\frac 52^-)$ states in the mass region $\la 2000$ MeV. We see that the
inclusion of the
 $SU(6)$ constraints on $qD^{1}_{1}$ states with angular momentum $L=1$
 results in a shift of the lightest $\Delta(\frac 52^-)$ isobar to
 $\sim 2300$ MeV. The scheme with the $SU(6)$ constraints for low-lying
 $qD^{1}_{1}$ and $qD^{0}_{0}$ states (with $L=0,1$) is presented in detail here.
 \end{abstract}

\ccode{PACS numbers: 11.25.Hf, 123.1K}

\section{Introduction}

Baryons (we mean standard non-exotic baryons) are composite systems of
three constituent quarks, each of them in their turn  being a
complicated system of quarks and gluons. We do not know the detailed
structure of baryons, even in the language of constituent quarks,
except for a fragmentary knowledge on low-lying states.

We know that low-lying baryons satisfy the $SU(6)$ symmetry, but as
to heavier ones we are not certain about. Numerous model
calculations, which describe rather well the low-lying states, are at
variance in their predictions concerning highly excited states, e.g.,
see \cite{Izgur-3q,Izgur-3qa,Izgur-3qb,Gloz,Petry} and references
therein. In addition, the number of highly excited states predicted by
such models exceeds considerably the number of observed states. One
might believe that experimental investigations of baryon spectra were
not complete. Nevertheless, it does not remove the question whether it
is possible to construct a more effective scheme for highly excited
states, with less degrees of freedom and less number of highly excited
states. The introduction of another effective particle, that is the
diquark, provides us with such a possibility.

The size of diquark as a composite quark--antiquark system is believed
to be of the order of that of quark, $\sim$0.2--0.3 fm
\cite{book2,book3}. So, it is doubtful if we can interpret the low--lying
baryon (a compact system, $\sim$0.7-0.9 fm) as a system of
effective diquark and constituent quark.

The idea of baryon as spatially separated quark and diquark can be for
sure tried on the highly excited baryon systems which are of a
larger size -- such an idea was suggested in \cite{qD}.
But in \cite{qD} we accepted also that the quark--diquark structure does not
work for low-lying baryons with  angular momentum
$L=0$ -- for these states the $SU(6)$ symmetry was applied.

Actually, in \cite{qD} we have made an attempt to systematize highly excited
baryons, assuming that they do not like to be formed as three-particle
colour quark systems but prefer to be created as two-particle,
quark--diquark, compound states:
$q_\alpha D^\alpha\equiv q_\alpha [\varepsilon^{\alpha\beta\gamma}
q_{\beta\gamma}]$
where $\varepsilon^{\alpha\beta\gamma}$ is a totally antisymmetrical tensor
in the colour space.

In \cite{qD} we have considered two  schemes of the
quark-diquark construction for the $L\ge 1$ states, namely:
 \\ (1) The
diquark masses $M_{D^0_0}\ne M_{D^1_1}$.
 \\ (2) The diquark masses are
equal to each other $M_{D^0_0}= M_{D^1_1}$; besides, the states
$qD^0_0$ and $qD^1_1$ with the total spin $S=1/2$ (here
${\bf S}={\bf S}_q+{\bf S}_{D}$) overlap. This scheme
decreases essentially the number of highly excited states.

In the quark--diquark scheme of \cite{qD} we face a specific transition to the
$SU(6)$ limit. This procedure is in fact a projection  of the $qD^0_0$
and $qD^1_1$ wave functions on the $SU(6)$ basis -- the result depends on the
hypothesis on diquark masses. In the $L=0$ states, at
$M_{D^0_0}= M_{D^1_1}$, we have two basic ($n=1$) states,
$N_{\frac 12^+}(940)$ and $\Delta_{\frac 32^+}(1240)$, see
equation (46) in \cite{qD}, while at $M_{D^0_0}\ne M_{D^1_1}$ the
additional basic state, $N_{\frac 12^+}(1440)$, appears  (see equation (49) in \cite{qD}).

\begin{figure}[h]
%Fig. 1
\centerline{\epsfig{file=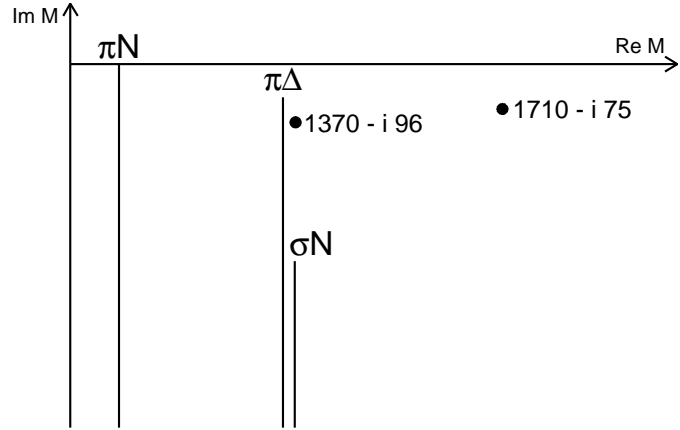, width=0.7\textwidth}}
\caption{\label{Vroper} Complex-$M$ plane:
position of     poles of resonance states $N(1440)$ and $N(1710)$.
The cuts related to the threshold singularities $\pi N$,
 $\pi\Delta$, $\rho N$ and
$\sigma N$ are
 shown by  vertical solid lines. The Roper pole is located near the
 $\pi\Delta$ cut, while its satellite pole is hidden under the cut.
 \label{Roper}} \end{figure}

In the present paper we accept the equality of masses of
scalar and axial--vector diquarks, $M_{D^0_0}= M_{D^1_1}$. In recent
analyses of spectra near the Roper resonance one may find arguments in favour of
the existence of one pole in the partial wave $P_{11}$ near 1400 MeV.
Analytical properties of the partial wave $P_{11}$ in the best fit
of \cite{An,Al} are shown in Fig. \ref{Roper}; the fitting to two
poles near 1400 MeV gives us a worse description of data. Of course, the
Roper pole splits, owing to the momentum-dependence of width near the
$\pi\Delta$ branching point. This splitting is quite similar to that
observed in the Flatt\'e formula \cite{flatte} for $f_0(980)$ near the
$K\bar K$ threshold. Still, we attribute such "satellite poles" to
the main one, in case in question, to the Roper pole ($1370\pm i96$);
note that in Fig. \ref{Roper} the satellite pole is hidden under the
$\pi\Delta$ cut.

There is a question to what low-lying states the $SU(6)$ symmetry may
be applied and where we have the region with spatially separated
quark and diquark. Only the experiment can answer this question. If
we turn to the PDG compilation \cite{PDG}, it may seem that the $SU(6)$
symmetry can be applied to $L=0$ only, while at $L\ge 1$ the domain of
quark--diquark structures begin. However, the latest analyses
\cite{An,Al,Arndt} give rise to doubts. The matter is that in the
experiments carried out in the eighties
\cite{Cutkosky:1980rh,Manley:1992yb,Hohler:1973ww} the resonance
$\Delta_{5/2^-}(1930\pm 50)$ has been observed rather definitely. Still,
 modern analyses \cite{An,Al,Arndt} point to the lower
$\Delta_{5/2^-}$ state, being located around  2200 MeV or even higher.
It gives arguments for  expanding  the $SU(6)$ symmetry constraints
on the $L=1$ states.

In the present paper we accept that both
sets of states  ($L=0$ and $L=1$) satisfy the $SU(6)$ symmetry
requirements.

The paper is organized as follows:\\
In Section 2 we present wave functions of quark--diquark states and
demonstrate how the imposing of the $SU(6)$ constraints affect these systems. In
Section 3 we suggest the setting of the $L=0$ and $L=1$ states, while
states with $L>1$ are discussed in Section 4. Here we demonstrate the setting of all
quark--diquark states on the ($J,M^2$)  and
($n,M^2$) trajectories.

\section{Wave functions of quark--diquark systems and the $SU(6)$
constraints   } \label{Section-WF-L123}

We present here the diquark wave function, give  general form for the
quark--diquark wave function and present the scheme of projecting them
on the $SU(6)$ basis.

\subsection{$S$-wave diquarks and baryons}

Recall that we use two $S$-wave diquarks with color numbers
${\bf\bar c}=3$: scalar diquark $ D^{0}_{0}$ and axial--vector  one,
$D^{1I_Z}_{1S_Z}$. The diquark spin--flavor wave functions
with $I_D=~1,\, S_D=1$ and with $I_D=0,\,S_D=0$ read as follows:
\bea \label{D-1}
&&
D^{11}_{11}(ij)= u^\uparrow (i)u^\uparrow (j),\nn \\
 &&D^{11}_{10}(ij)= \frac{1}{\sqrt 2}
 \bigg(u^\uparrow (i)u^\downarrow (j)+ u^\downarrow (i)u^\uparrow
(i)\bigg),\nn \\
&&D^{10}_{11}(ij)= \frac{1}{\sqrt 2}\bigg(u^\uparrow
(i)d^\uparrow (j)+ d^\uparrow (i)u^\uparrow (j)\bigg),\nn\\
&&D^{10}_{10}(ij)= \frac{1}{ 2}\bigg(u^\uparrow (i)d^\downarrow (j)+
u^\downarrow (i)d^\uparrow (j)+d^\uparrow (i)u^\downarrow (j)+
d^\downarrow (i)u^\uparrow (j)\bigg),\nn \\
&& D^{0}_{0}(ij)= \frac{1}{ 2}\bigg(u^\uparrow (i)d^\downarrow (j)-
u^\downarrow (i)d^\uparrow (j)-d^\uparrow (i)u^\downarrow (j)+
d^\downarrow (i)u^\uparrow (j)\bigg).
\eea

Let us consider, first, the $\Delta$ isobar at $I_Z=3/2$ with fixed
$J,J_Z$,  total spin $S$ and orbital momentum $L$.
The wave function for this state at arbitrary $n$ reads
\bea \label{L-28}
&&
\sum\limits_{S_Z,m_z}
 C^{J\,J_Z}_{L\,J_Z-S_Z\;\;S\,S_Z}
 C^{S\,S_Z}_{1\,S_Z-m_z\;\;\frac 12\,m_z}
 C^{\frac 32\frac 32}_{1\,1\;\;\frac 12\frac 12}
 \Big(  u^{m_z} (1)D^{11}_{1\, S_Z-m_z}(23)
\nn
\\
&&\qquad\qquad\times
|\vec k_{1\, cm}|^LY_{L}^{J_Z-S_Z}(\theta_1,\phi_1)\Phi_1^{(L)}(1;23)
+(1\rightleftharpoons 2)
  +(1\rightleftharpoons 3)
      \Big)\, .
  \eea
Here $|\vec k_{1\, cm}|$  and $(\theta_1,\phi_1)$ are the momenta
and momentum angles of the first quark in the c.m. system.

For other $I_Z$, one should include into the wave function a summing over
isotopic states which means the following substitution in (\ref{L-28}):
\beq \label{L-29}
 C^{\frac 32\frac 32}_{1\,1\;\;\frac 12\frac 12}
  u^{m_z} (1)D^{11}_{1\, S_Z-m_z}(23)\to
\sum\limits_{j_z}
 C^{\frac 32 I_Z}_{1\,I_Z-j_z\;\;\frac 12 \, j_z}
  q^{m_z}_{j_z} (1)D^{1\, I_Z-j_z}_{1\, S_Z-m_z}(23)  .
  \eeq
To project the wave functions (\ref{L-28}) or
(\ref{L-29}) on the $SU(6)$ wave function set, we should use symmetrical
coordinate/momentum wave functions
\beq
\Phi_{I_D}^{(L)}(\ell;ij)\to \Phi_{I_D}^{(L)}(\ell,i,j) =
\Phi_{I_D}^{(L)}(j,\ell,i)=\Phi_{I_D}^{(L)}(i,j,\ell).
\eeq
For example, one may accept that
 $\Phi_{I_D}^{(L)}(\ell,i,j)$
depends on $s$
only, $\Phi_{I_D}^{(L)}(\ell,i,j)\to  \varphi_{I_D}^{(L)}(s)$.
 Indeed, in this limit we have
\bea \label{L-30}
&&
\sum\limits_{S_Z,m_z}
 C^{J\,J_Z}_{L\,J_Z-S_Z\;\;S\,S_Z}
 C^{S\,S_Z}_{1\,S_Z-m_z\;\;\frac 12\,m_z}
\sum\limits_{j_z}
 C^{\frac 32 I_Z}_{1\,I_Z-j_z\;\;\frac 12 \, j_z}
\nn \\
&&\times\Big(  q^{m_z}_{j_z} (1)D^{1\, I_Z-j_z}_{1\, S_Z-m_z}(23)
|\vec k_{1\, cm}|^LY_{L}^{J_Z-S_Z}(\theta_1,\phi_1)
+(1\rightleftharpoons 2)
  +(1\rightleftharpoons 3)
      \Big)\varphi_1^{(L)}(s)\, .\qquad
  \eea

For nucleon states $(I=1/2)$ we write
\bea \label{L-31}
&&
\sum\limits_{S_Z,m_z}
 C^{J\,J_Z}_{L\,J_Z-S_Z\;\;S\,S_Z}
 C^{S\,S_Z}_{1\,S_Z-m_z\;\;\frac 12\,m_z}
\sum\limits_{j_z}
 C^{\frac 12 I_Z}_{1\,I_Z-j_z\;\;\frac 12 \, j_z}
 \Big(  q^{m_z}_{j_z} (1)D^{1\, I_Z-j_z}_{1\, S_Z-m_z}(23)
\nn \\
&&\qquad\qquad\times
|\vec k_{1\, cm}|^LY_{L}^{J_Z-S_Z}(\theta_1,\phi_1)
\Phi_{1}^{(L)}(1;23) +(1\rightleftharpoons 2)
  +(1\rightleftharpoons 3)
      \Big)\, .\qquad
  \eea

The $SU(6)$ limit, as previously,
is reached at $\Phi_{1}^{(L)}(i;j\ell)\to\varphi_1^{(L)}(s)$. Then,
instead of  (\ref{L-31}), one has
\bea \label{L-31a}
&&
\sum\limits_{S_Z,m_z}
 C^{J\,J_Z}_{L\,J_Z-S_Z\;\;S\,S_Z}
 C^{S\,S_Z}_{1\,S_Z-m_z\;\;\frac 12\,m_z}
\sum\limits_{j_z}
 C^{\frac 12 I_Z}_{1\,I_Z-j_z\;\;\frac 12 \, j_z}
\nn \\
&&\times\Big(  q^{m_z}_{j_z} (1)D^{1\, I_Z-j_z}_{1\, S_Z-m_z}(23)
|\vec k_{1\, cm}|^LY_{L}^{J_Z-S_Z}(\theta_1,\phi_1)
 +(1\rightleftharpoons 2)
  +(1\rightleftharpoons 3)
      \Big)\varphi_{1}^{(L)}(s)\, .\qquad
  \eea
For $qD^0_0$ states the wave function reads in general case as follows:
\bea \label{L-31b}
&&
\sum\limits_{m_z}
 C^{J\,J_Z}_{L\,J_Z-m_z\;\;\frac 12\,m_z}
\Big(  q^{m_z}_{I_z} (1)D^{0}_{0}(23)
|\vec k_{1\, cm}|^LY_{L}^{J_Z-m_z}(\theta_1,\phi_1)\Phi_{0}^{(L)}(1;23)
\nn \\
&&\qquad\quad\qquad\qquad\qquad
+(1\rightleftharpoons 2)
  +(1\rightleftharpoons 3)
      \Big)\ . \qquad
  \eea
In the $SU(6)$ limit we have
\bea \label{L-31c}
&&
\sum\limits_{m_z}
 C^{J\,J_Z}_{L\,J_Z-m_z\;\;\frac 12\,m_z}
\Big(  q^{m_z}_{I_z} (1)D^{0}_{0}(23)
|\vec k_{1\, cm}|^LY_{L}^{J_Z-m_z}(\theta_1,\phi_1)
\nn \\
 &&\qquad\quad\qquad\qquad\qquad
 +(1\rightleftharpoons 2)
  +(1\rightleftharpoons 3)
      \Big)\varphi_0^{(L)}(s)\, .\qquad
  \eea
  Baryons are characterized by $I$ and $J^P$, these states with
  different $S$ and $L$ and fixed $I$ and $J^P$ can mix. To select independent states, one
  may orthogonalize wave functions with the same isospin and $J^P$. The
  orthogonalization depends on the structure of the momentum/coordinate
  parts $\Phi_{1}^{(L)}(i;j\ell)$. But in  case of the $SU(6)$ limit
  the momentum/coordinate wave functions transform in the common
  factor $\Phi_{0}^{(L)}(i;j\ell)\to\varphi_{0}^{(L)}(i,j,\ell)$ or
   $\Phi_{1}^{(L)}(i;j\ell)\to\varphi_{SU(6)}^{(L)}(s)$, and one should
   orthogonalize the spin/momentum
  factors. Namely, let us denote the $SU(6)$ spin/momentum factor of
  the wave function as $Q^{(A)}_{J^P}$. Then
\beq \label{L-31d}
  \Psi^{(A)}_{J^P}=Q^{(A)}_{J^P}\varphi_{SU(6)}^{(A)}(s)\ ,
\eeq
where
$A=I,II,III,...$ belong to different $(S,L)$. The orthogonal set of operators
$Q^{(A)}_{J^P}$ is constructed in a standard way:
\bea \label{L-31e}
&& Q^{(\perp I)}_{J^P}\equiv Q^{(I)}_{J^P}\, ,
\nn \\
&&
Q^{(\perp II)}_{J^P}= Q^{( II)}_{J^P}-Q^{(\perp I)}_{J^P}
\frac{\Big(Q^{(\perp I)+}_{J^P}Q^{( II)}_{J^P} \Big)}
{\Big(Q^{(\perp I)+}_{J^P}Q^{(\perp I)}_{J^P} \Big)}\, ,
\nn \\
&&
Q^{(\perp III)}_{J^P}= Q^{(III)}_{J^P}-
Q^{(\perp I)}_{J^P}
\frac{\Big(Q^{(\perp I)+}_{J^P}Q^{( III)}_{J^P} \Big)}
{\Big(Q^{(\perp I)+}_{J^P}Q^{(\perp I)}_{J^P} \Big)}-
Q^{(\perp II)}_{J^P}
\frac{\Big(Q^{(\perp II)+}_{J^P}Q^{( III)}_{J^P} \Big)}
{\Big(Q^{(\perp II)+}_{J^P}Q^{(\perp II)}_{J^P} \Big)},
  \eea
and so on. The convolution of operators
$\Big(Q^{(A)+}_{J^P}Q^{(B)}_{J^P} \Big)$ includes  the summation over
quark spins as well as integration over quark momenta.

\subsection{Wave functions in the $SU(6)$ limit} \label{WF-L1-SU6-limit}

We present wave functions of the $qD^0_0$ and $qD^1_1$ systems in the
$SU(6)$ limit, assuming $M^2_{D^0_0}= M^2_{D^1_1}\equiv  M^2$. The
 transition $M^2_{D^0_0}\to M^2_{D^1_1}$ gives us additional
reduction of number of states.

Recall that in the $SU(6)$ limit the momentum/coordinate factors of wave functions, for
example, such as $\Phi^{(L)}_0 (3,12)=\Phi^{(L)}_0 (s,s_{12})$ and
$\Phi^{(L)}_1 (3,12)=\Phi^{(L)}_1 (s,s_{12})$,  transform into $s$-dependent ones:
\bea \label{wf-su3}
 &&\Phi^{(L)}_0 (s,s_{12}\to
M^2_{D^0_0} )= \varphi^{(L)}_0(s,M^2_{D}) \equiv
\varphi^{(L)}_0(s),  \\
&&\Phi^{(L)}_1 (s,s_{12}\to M^2_{D^1_1})=
\varphi^{(L)}_1(s,M^2_{D}) \equiv \varphi^{(L)}_1(s). \nn
\eea
For
our purpose, that is, for checking the number of non-vanishing states, we
calculate below the wave functions with $I=I_z$ and $J=J_z$ only.

\section{Wave functions of the ($L=1$) states
in the $SU(6)$ limit and $M^2_{D^0_0}\to M^2_{D^1_1}$
}

In this sector we have only seven basic states, namely,\\
(1) two with $I=1/2$:
$\qquad N_{\frac 12^-}(D^0_0)=N_{\frac 12^-}(D^1_1;S=1/2)$,\\ \qquad
$N_{\frac 32^-}(D^0_0)=N_{\frac 32^-}(D^1_1;S=1/2)$,  \\
(2) three  with $I=1/2$: $\qquad N_{\frac 12^-}(D^1_1;S=3/2)$,
$\quad N_{\frac 32^-}(D^1_1;S=3/2)$,\\ $\qquad N_{\frac
52^-}(D^1_1;S=3/2)$,\\ (3) two with  $I=3/2$: $\qquad \Delta_{\frac
12^-}(D^1_1;S=1/2)$, $\quad \Delta_{\frac 32^-} (D^1_1;S=1/2)$. \\

 Below the wave functions are written in c.m. system
using the following notations for quark momenta: $\vec k_1+\vec
k_2+\vec k_3=0$, $ k_{a\pm}=(k_{ax}\pm ik_{ay})/\sqrt2$.

{\bf (1) Two $qD^0_0$ systems with $I=1/2$}

We have two basic states:
\bea \label{AA1}
&&4\sqrt{\pi}\cdot \Psi^{(L=1)}_{0}\Big(I=
\frac12,\;I_z=\frac12,\; J^P=\frac12 ^-,\; J_z=\frac12\Big)
\nn \\
&=&
\Big(
  u_1\up u_2\up d_3\dw k_{3z}
 + u_1\up u_2\dw d_3\up \kaz
 -\sqrt2 u_1\up u_2\dw d_3\dw \kbp
 + u_1\up d_2\up u_3\dw \kaz  \Big .
+ u_1\up d_2\dw u_3\up \kbz
 \nn\\
&&
 -\sqrt2 u_1\up d_2\dw u_3\dw k_{3+}
 + u_1\dw u_2\up d_3\up \kbz
 -\sqrt2 u_1\dw u_2\up d_3\dw \kap
-\sqrt2 u_1\dw u_2\dw d_3\up k_{3+}
 \nn\\
&&
 + u_1\dw d_2\up u_3\up k_{3z}
 -\sqrt2 u_1\dw d_2\up u_3\dw \kbp
 -\sqrt2 u_1\dw d_2\dw u_3\up \kap
 + d_1\up u_2\up u_3\dw \kbz
+ d_1\up u_2\dw u_3\up k_{3z}
    \nn\\
&&
\Big .
 -\sqrt2 d_1\up u_2\dw u_3\dw \kap
 + d_1\dw u_2\up u_3\up \kaz
 -\sqrt2 d_1\dw u_2\up u_3\dw k_{3+}
 -\sqrt2 d_1\dw u_2\dw u_3\up \kbp  \Big )\varphi_0^{(L=1)}(s)
 \qquad
\eea
and
\bea \label{AA2}
&&4\sqrt{\pi}\cdot \Psi^{(L=1)}_{0}\Big(I=
\frac12,\;I_z=\frac12,\; J^P=\frac32 ^-,\; J_z=\frac32\Big)
\nn \\
&=&
\sqrt{3}\Big(
 u_1\up u_2\up d_3\dw k_{3+}
 + u_1\up u_2\dw d_3\up \kap
 + u_1\up d_2\up u_3\dw \kap
 + u_1\up d_2\dw u_3\up \kbp
 + u_1\dw u_2\up d_3\up \kbp      \Big.       \nn \\
 &&
 \Big.
 + u_1\dw d_2\up u_3\up k_{3+}
 + d_1\up u_2\up u_3\dw \kbp
 + d_1\up u_2\dw u_3\up k_{3+}
 + d_1\dw u_2\up u_3\up \kap    \Big )\varphi_0^{(L=1)}(s).
 \qquad
\eea

The same angular momentum/spin wave functions can be constructed with
use of the $qD^1_1$ system, in this way
we have two nucleons with $S=1/2$:
\bea \label{AA3}
&&4\sqrt{\pi}\cdot \Psi^{(L=1)}_{1}\Big(I=
\frac12,\;I_z=\frac12,\;S=1/2,\; J^P=\frac12 ^-,\; J_z=\frac12\Big)
\nn \\
&=&
-\sqrt{2}\Big(
  u_1\up u_2\up d_3\dw k_{3z}
 + u_1\up u_2\dw d_3\up \kaz
 -\sqrt{2} u_1\up u_2\dw d_3\dw \kbp
 + u_1\up d_2\up u_3\dw \kaz  \Big .
+ u_1\up d_2\dw u_3\up \kbz
 \nn\\
&&
 -\sqrt{2} u_1\up d_2\dw u_3\dw k_{3+}
 + u_1\dw u_2\up d_3\up \kbz
 -\sqrt{2} u_1\dw u_2\up d_3\dw \kap
-\sqrt{2} u_1\dw u_2\dw d_3\up k_{3+}
 + u_1\dw d_2\up u_3\up k_{3z}
 \nn\\
&&
 -\sqrt{2} u_1\dw d_2\up u_3\dw \kbp
 -\sqrt{2} u_1\dw d_2\dw u_3\up \kap
 + d_1\up u_2\up u_3\dw \kbz
+ d_1\up u_2\dw u_3\up k_{3z}
    \nn\\
&&
\Big .
 -\sqrt{2} d_1\up u_2\dw u_3\dw \kap
 + d_1\dw u_2\up u_3\up \kaz
 -\sqrt{2} d_1\dw u_2\up u_3\dw k_{3+}
 -\sqrt{2} d_1\dw u_2\dw u_3\up \kbp  \Big )\varphi_1^{(L=1)}(s)
 \qquad
\eea
and
\bea \label{AA4}
&&4\sqrt{\pi}\cdot \Psi^{(L=1)}_{1}\Big(I=
\frac12,\;I_z=\frac12,\;S=1/2,\; J^P=\frac32 ^-,\; J_z=\frac32\Big)=
\nn \\
&=&
\sqrt{3}\Big(
 u_1\up u_2\up d_3\dw k_{3+}
 + u_1\up u_2\dw d_3\up \kap
 + u_1\up d_2\up u_3\dw \kap
 + u_1\up d_2\dw u_3\up \kbp
 + u_1\dw u_2\up d_3\up \kbp      \Big.       \nn \\
 &&
 \Big.
 + u_1\dw d_2\up u_3\up k_{3+}
 + d_1\up u_2\up u_3\dw \kbp
 + d_1\up u_2\dw u_3\up k_{3+}
 + d_1\dw u_2\up u_3\up \kap    \Big )\varphi_1^{(L=1)}(s).
\eea
The wave function of (\ref{AA1}) coincides with that of (\ref{AA3}),
while the wave function of (\ref{AA2}) coincides with (\ref{AA4}), because we
use equation (\ref{wf-su3}):
$\varphi_0^{(L=1)}(1,2,3)=\varphi_1^{(L=1)}(1,2,3)$.
 In this case we
deal with two (not four) states.

{\bf (2) Three $qD^1_1$ systems with $I=1/2$}

 There are three states with $S=3/2$:

%$I_z=\frac12,\; J=\frac12,\; J_z=\frac12,\; S=\frac32$:
\bea \label{AA5a}
&&4\sqrt{\pi}\cdot \Psi^{(L=1)}_{1}\Big(I=
\frac12,\;I_z=\frac12,\;S=3/2,\; J^P=\frac12 ^-,\; J_z=\frac12\Big)
\nn \\
&=&
\Big(
 3 u_1\up u_2\up d_3\up        k_{3-}
 -\sqrt{2} u_1\up u_2\up d_3\dw k_{3z}
 -\sqrt{2} u_1\up u_2\dw d_3\up k_{3z}
 -u_1\up u_2\dw d_3\dw          k_{3+} \Big.
 +3 u_1\up d_2\up u_3\up \kbm
 \nn \\ &&
 -\sqrt{2} u_1\up d_2\up u_3\dw \kbz
 -\sqrt{2} u_1\up d_2\dw u_3\up \kbz
 -u_1\up d_2\dw u_3\dw \kbp
 -\sqrt{2} u_1\dw u_2\up d_3\up  k_{3z}
 \nn \\ &&
 -u_1\dw u_2\up d_3\dw           k_{3+}
 -u_1\dw u_2\dw d_3\up           k_{3+}
 -\sqrt{2} u_1\dw d_2\up u_3\up \kbz
-u_1\dw d_2\up u_3\dw \kbp
 \nn \\  &&
 -u_1\dw d_2\dw u_3\up \kbp
 +3 d_1\up u_2\up u_3\up \kam
 -\sqrt{2} d_1\up u_2\up u_3\dw \kaz
 -\sqrt{2} d_1\up u_2\dw u_3\up \kaz
 \nn \\ &&
\Big.
  -d_1\up u_2\dw u_3\dw \kap
 -\sqrt{2} d_1\dw u_2\up u_3\up \kaz
 -d_1\dw u_2\up u_3\dw \kap
 -d_1\dw u_2\dw u_3\up \kap \Big)\varphi_1^{(L=1)}(s),
\eea

%$I_z=\frac12,\; J=\frac32,\; J_z=\frac32, S=\frac32$:
\bea \label{AA6a}
&&4\sqrt{\pi}\cdot \Psi^{(L=1)}_{1}\Big(I=
\frac12,\;I_z=\frac12,\;S=3/2,\; J^P=\frac32 ^-,\; J_z=\frac32\Big)
\nn \\
&=&
\frac {3}{\sqrt{15}} \Big(
 -3 \sqrt{2} u_1\up u_2\up d_3\up  k_{3z}
 -2 u_1\up u_2\up d_3\dw  k_{3+}
 -2 u_1\up u_2\dw d_3\up  k_{3+}
 -3 \sqrt{2} u_1\up d_2\up u_3\up \kbz\Big.  \nn \\
       &&
 -2 u_1\up d_2\up u_3\dw \kbp
 -2 u_1\up d_2\dw u_3\up \kbp
 -2 u_1\dw u_2\up d_3\up  k_{3+}
 -2 u_1\dw d_2\up u_3\up \kbp  \nn \\
      &&
\Big.
 -3 \sqrt{2} d_1\up u_2\up u_3\up \kaz
 -2 d_1\up u_2\up u_3\dw \kap
 -2 d_1\up u_2\dw u_3\up \kap
 -2 d_1\dw u_2\up u_3\up \kap\;\Big)\varphi_1^{(L=1)}(s) ,\qquad
\eea
and
%$I_z=\frac12,\; J=\frac52,\; J_z=\frac52, S=\frac32$:
\bea \label{AA7a}
&&4\sqrt{\pi}\cdot \Psi^{(L=1)}_{1}\Big(I=
\frac12,\;I_z=\frac12,\;S=3/2,\; J^P=\frac52 ^-,\; J_z=\frac52\Big)
\nn \\
&=&
3\sqrt{2}\Big(
 - u_1\up u_2\up d_3\up k_{3+}
 - u_1\up d_2\up u_3\up \kbp
 - d_1\up u_2\up u_3\up \kap \Big)
\varphi_1^{(L=1)}(s).
\eea

{\bf (3) Two $qD^1_1$ states with $I=3/2$  }

%\subsubsection{ $\Delta$ isobars with $S=\frac12$}

In the ($I=\frac 32 ,\;L=1$) sector of  $qD^1_1$ states, there are
two baryons only in the $SU(6)$ limit -- the states with $S=\frac12$:
%$I_z=\frac32,\; J=\frac12,\; J_z=\frac12,\; S=\frac12$:
\bea \label{AA8}
&&4\sqrt{\pi}\cdot \Psi^{(L=1)}_{1}\Big(I=
\frac32,\;I_z=\frac32,\;S=1/2,\; J^P=\frac12 ^-,\; J_z=\frac12\Big)
\nn \\
&=&
\Big(
 -\sqrt6  u_1\up u_2\up u_3\dw k_{3z}
 -\sqrt6  u_1\up u_2\dw u_3\up \kbz
 -\sqrt6  u_1\dw u_2\up u_3\up \kaz \Big. \nn \\
&&
\Big.
+2\sqrt3 u_1\up u_2\dw u_3\dw \kap
 +2\sqrt3 u_1\dw u_2\up u_3\dw \kbp
 +2\sqrt3 u_1\dw u_2\dw u_3\up k_{3+}\Big)\varphi_1^{(L=1)}(s) ,
\eea
and
\bea \label{AA9}
&&4\sqrt{\pi}\cdot
\Psi^{(L=1)}_{1}\Big(I=
\frac32,\;I_z=\frac32,\;S=1/2,\; J^P=\frac32 ^-,\; J_z=\frac32\Big)
\nn \\
&=&
-3\sqrt{2}\Big(
  u_1\up u_2\up u_3\dw k_{3+}
 + u_1\up u_2\dw u_3\up \kbp
 + u_1\dw u_2\up u_3\up \kap\Big)\varphi_1^{(L=1)}(s).
\eea

\subsection{The vanishing class of ($S=3/2, \,I=3/2$) states}

In  the $SU(6)$ limit, the states with $S=\frac32,\; I=\frac 32 $
(i.e. $\Delta_{\frac 12^-}$, $\Delta_{\frac 32^-}$, $\Delta_{\frac 52^-}$) are absent. It
is easy to be certain of that, after calculating wave functions with $J_z=J$. Indeed,
\bea \label{AA10} && \Psi^{(L=1)}_{1}\Big(I= \frac32,\;I_z=\frac32,\;S=3/2,\; J^P=\frac12
^-,\; J_z=\frac12\Big)= 0,\nn \\ && \Psi^{(L=1)}_{1}\Big(I=
\frac32,\;I_z=\frac32,\;S=3/2,\; J^P=\frac32 ^-,\; J_z=\frac32\Big)= 0,\nn\\ &&
\Psi^{(L=1)}_{1}\Big(I= \frac32,\;I_z=\frac32,\;S=3/2,\; J^P=\frac52 ^-,\;
J_z=\frac52\Big)= 0. \eea
The absence of  $\Delta_{\frac 52^-}$, which, being in the $L=1$ set, should be
located at $\sim 1900$ MeV, is a primary motivation for expanding
 the $SU(6)$ symmetry onto  $L=1$ states.

\section{The $SU(6)$ symmetry and the $L=0$ and $L=1$
states}

First, we recall the setting of $L=0$ states,
then we discuss the situation with $L=1$ states under the $SU(6)$ symmetry.

\subsection{The $SU(6)$ symmetry for the nucleon $N_{\frac 12^+}(940)$,
 isobar  $\Delta_{\frac 32 ^+}(1238)$ and their radial
excitations }

We accept  $M^2_{D^0_0}=M^2_{D^1_1}$ and
suppose the $SU(6)$ symmetry for the lowest baryons with $L=0$. It gives
us two ground states -- the nucleon $N_{\frac 12^+}(940)$  and isobar
$\Delta_{\frac 32 ^+}(1238)$ as well as their radial excitations:
\begin{equation}\label{L01}
{\renewcommand{\arraystretch}{0,5}
\begin{tabular}{l|ll}
 $L=0$         &$S=\frac12$, $N(\frac12^+)$&
$S=\frac32$, $\Delta(\frac 32^+) $ \\
      $n=1$    &$ 938\pm 2 $&$1232\pm 4$  \\
      $n=2$    &$1440\pm 40 $&$1635\pm 75$ \\
      $n=3$    &$1710\pm 30 $&$\sim 1880  $\\
      $n=4$    &$2090\pm 100$&$\sim 2150  $  \\
\end{tabular}
}
\end{equation}
Note that the mass-squared splitting of the nucleon
radial excitation  states, $\delta_n M^2(N_{\frac 12^+})$, is of the
order of $1.05\pm 0.15$ GeV$^2$.
 This value is close to that observed in
 meson sector \cite{book3,syst}:
\bea
\label{18spl}
&&M^2[N_{\frac 12^+}(1440)]-M^2[N_{\frac 12^+}(940)]
\nn \\ \simeq
&&M^2[N_{\frac 12^+}(1710)]-M^2[N_{\frac 12^+}(1440)]\equiv
\delta_n M^2(N_{\frac 12^+})\simeq 1.0 {\rm GeV}^2.
\eea
 The state with $n=4$ cannot be unambiguously determined.

One can see that the mass-squared splitting of
$\Delta_{\frac 32^+}$ isobars,
$\delta_n M^2(\Delta_{\frac 32^+})$, coincides with
that of the  nucleon, $\delta_n M^2(N_{\frac 12^+})$, with a good
accuracy:
\beq \label{17b}
\delta_n M^2(\Delta_{\frac 32^+})=1.07\pm 0.05  . \eeq
Let us emphasize that state $ \Delta_{\frac 32^+}(1920)$ is classified
as $S=3/2,L=2$ states, with $n=1$ (see Section 4).
However, this resonance can be reliably classified as
radial excitations of $\Delta_{\frac 32^+}(1232)$, with $n=3$.
Actually, it means that
around $\sim 1920$ MeV  one may expect the double-pole structure.

\subsection{Setting of the $L=1$ states under the $SU(6)$ symmetry
constraints }

\begin{equation}
{\renewcommand{\arraystretch}{0,5}
\begin{tabular}{l|l|l|llll}
$L=1$&$S=\frac12$&     &    &$N(\frac12^-)$&$N(\frac32^-)$&   \\
     &     &$n=1$&    &$(1535\pm 20) $&$(1524\pm  5)$&  \\
     &     &$n=2$&    &$(1905\pm 60) $&$(1870\pm 25)$&  \\
     &     &$n=3$&    &$(2090\pm 150)$&$(2160\pm 35)$&  \\
     &     &$n=4$&    &$\sim 2390$&$\sim 2390$&  \\
&$S=\frac32$&     &$   $&$N(\frac12^-)$&$N(\frac32^-)$&$N(\frac52^-)$     \\
     &     &$n=1$&$         $&$(1680\pm 40)$&$(1730\pm 40)$&$(1680\pm 10)$  \\
     &     &$n=2$&$         $&$\sim 2010$&$\sim 2000$&$\sim 2000$  \\
     &     &$n=3$&$         $&$\sim 2270$&$\sim 2270$&$\sim 2260$  \\
     &     &$n=4$&$         $&$\sim 2500$&$\sim 2500$&$\sim 2500$  \\
\end{tabular}
}
\end{equation}

\begin{equation}
{\renewcommand{\arraystretch}{0,5}
\begin{tabular}{l|l|l|llll}
$L=1$&$S=\frac12$&     &    &$\Delta(\frac12^-)$&$\Delta(\frac32^-)$&   \\
     &     &$n=1$&    &$(1625\pm 10)$&$(1720\pm 50)$&  \\
     &     &$n=2$&    &$(1910\pm 50)$&$(1995\pm 40)$&  \\
     &     &$n=3$&    &$(2150\pm 50)$&$\sim 2210$&  \\
     &     &$n=4$&    &$\sim 2460$&$\sim 2450$&  \\
\end{tabular}
}
\end{equation}

\section{The setting of states with $L\ge 2$}

For a "naive observer",
who does not perform an analysis of double pole structure, the
number of states with $S=1/2$ decreases twice.

\begin{figure}[h]
%Fig. 2
\centerline{\epsfig{file=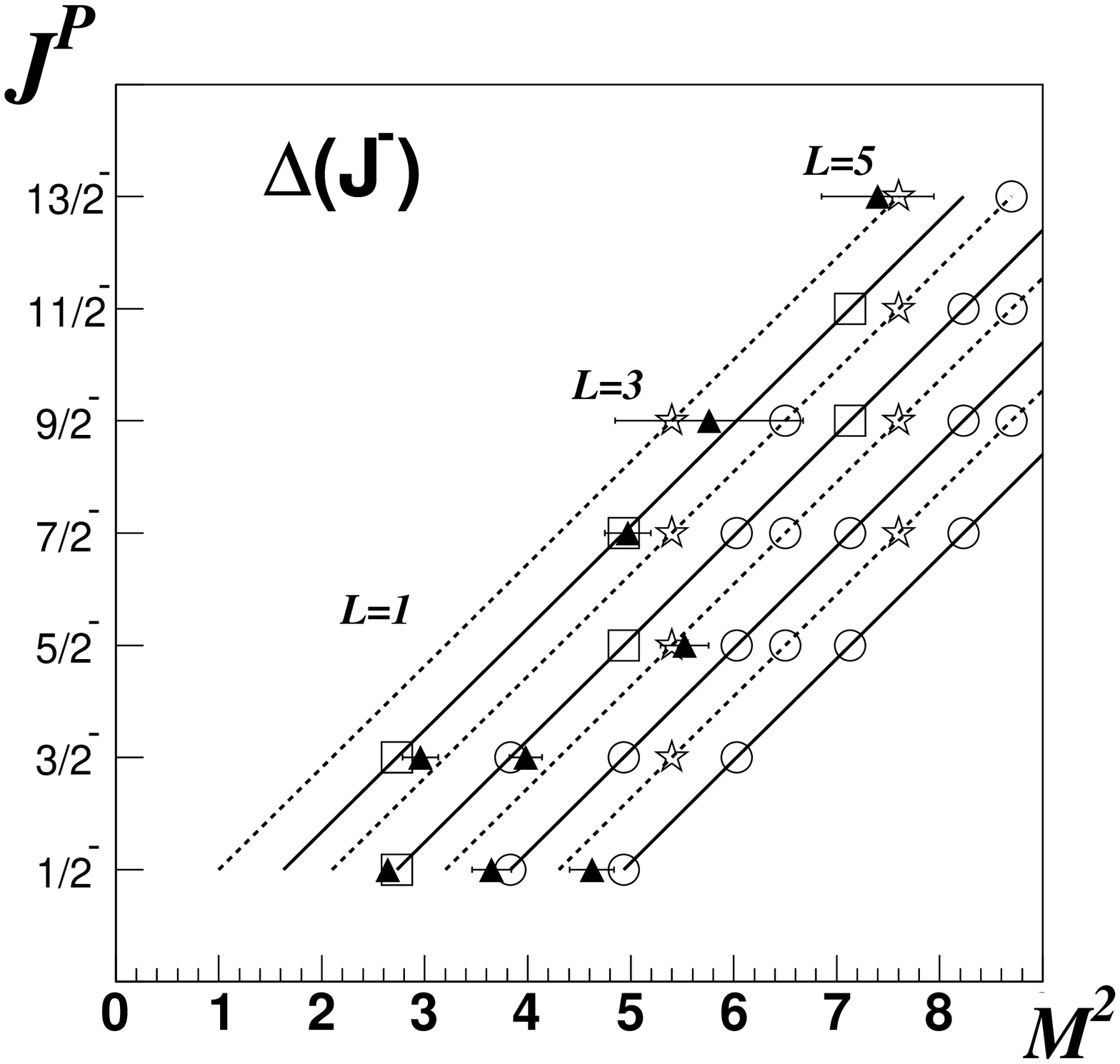,width=70mm}
             \epsfig{file=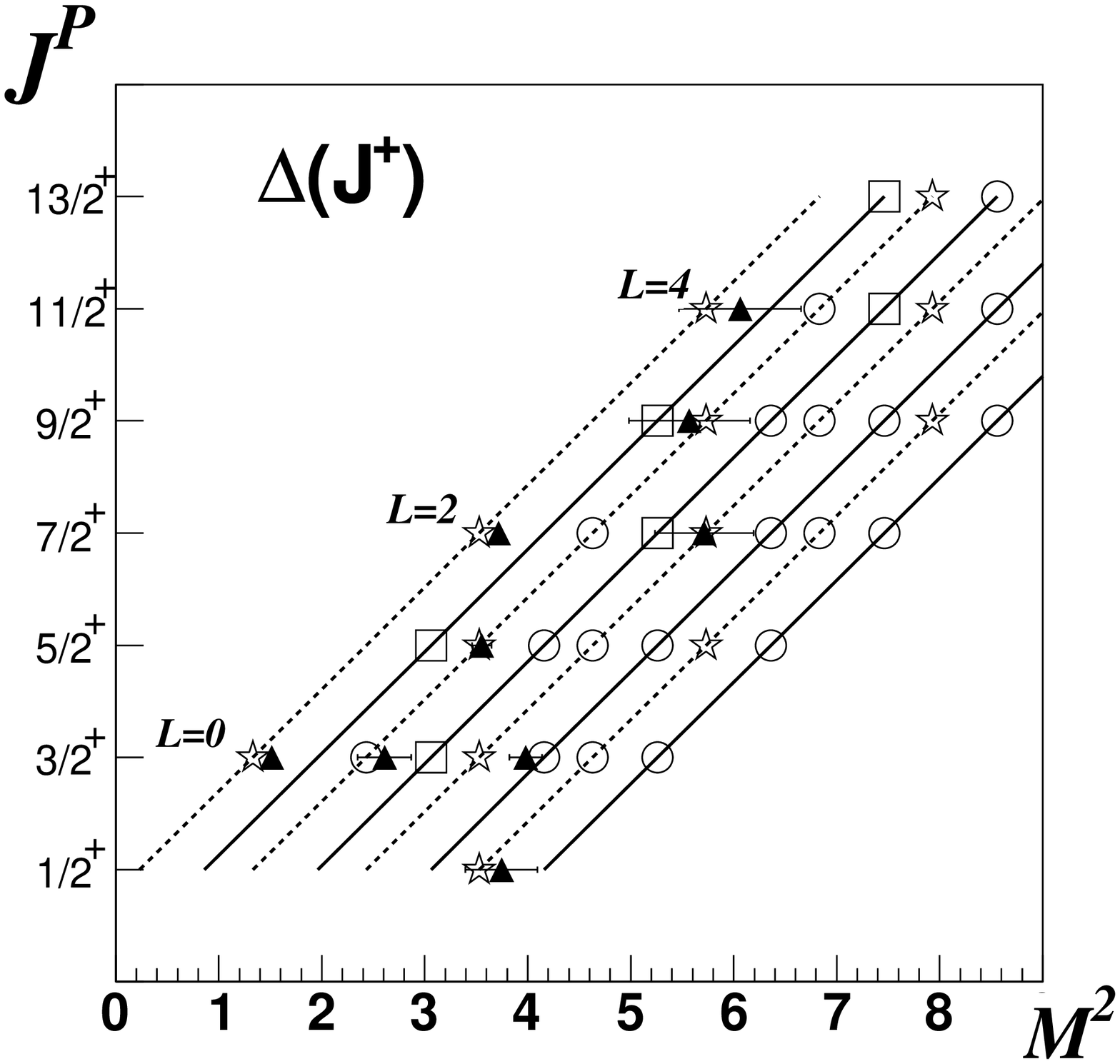,width=70mm}}
\centerline{\epsfig{file=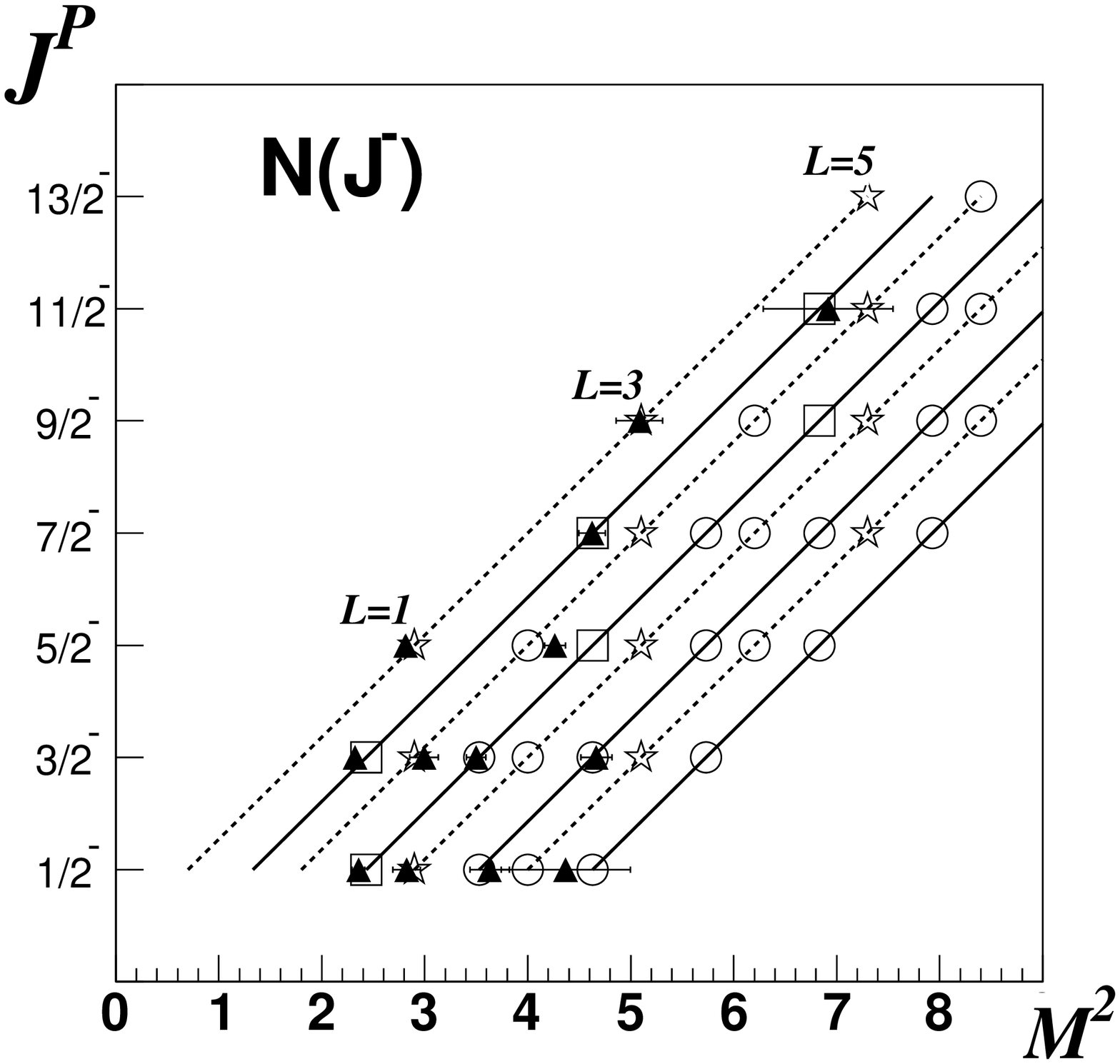,width=70mm}
             \epsfig{file=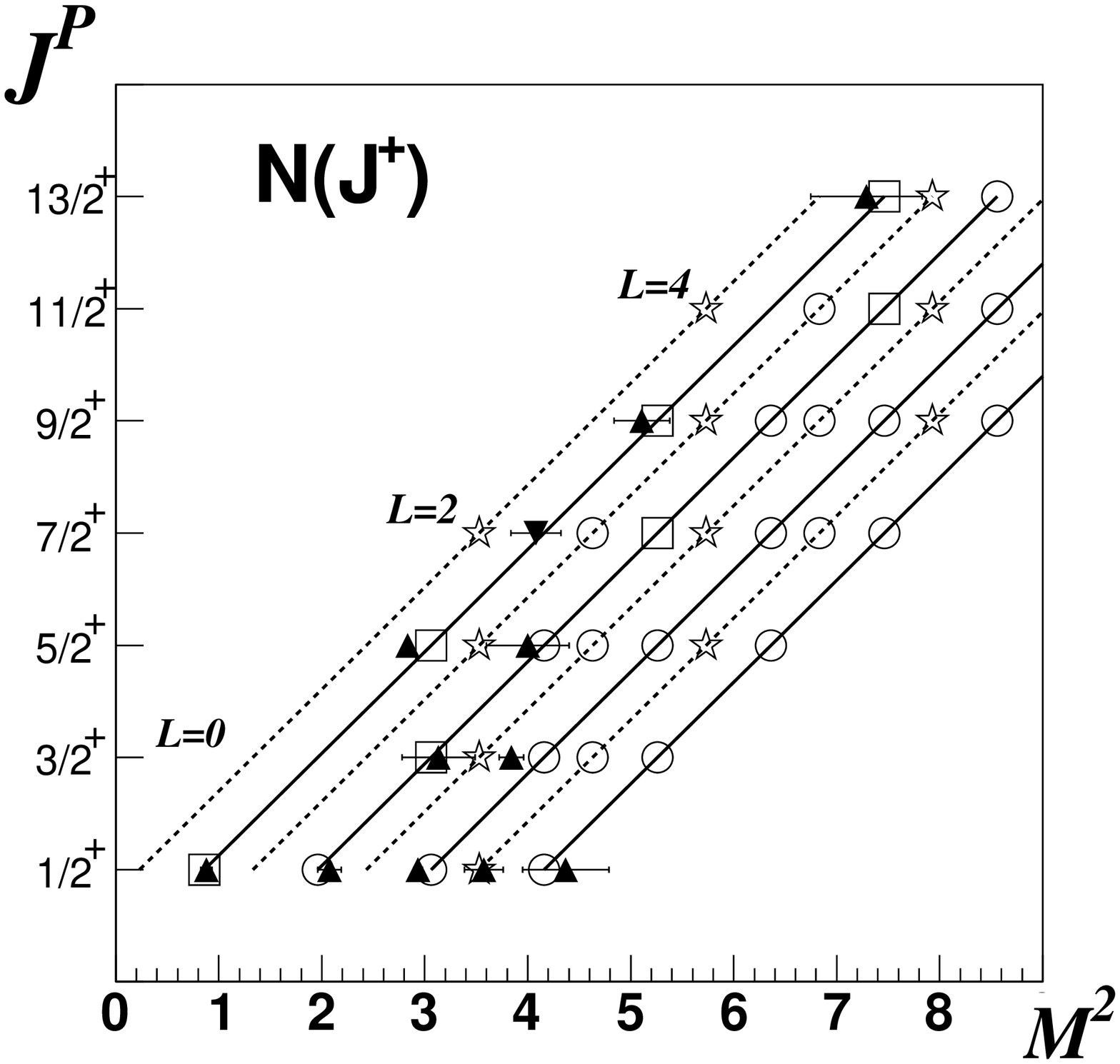,width=70mm}}
\caption{ Baryon settings  on
$(J^P,M^2)$ planes in the model with overlapping $qD^0_0(S=1/2)$ and
$qD^1_1(S=1/2)$ states. Notations are as follows:
1) open squares: predicted $S=1/2$ ground states,
2) open stars: predicted $S=3/2$ ground states,
3) open circles: predicted radial excitation states,
4) full triangles: observed states.}
%}
%}
\label{J-M-eq-M}
\end{figure}

\begin{figure}[h]
%Fig. 3
\centerline{\epsfig{file=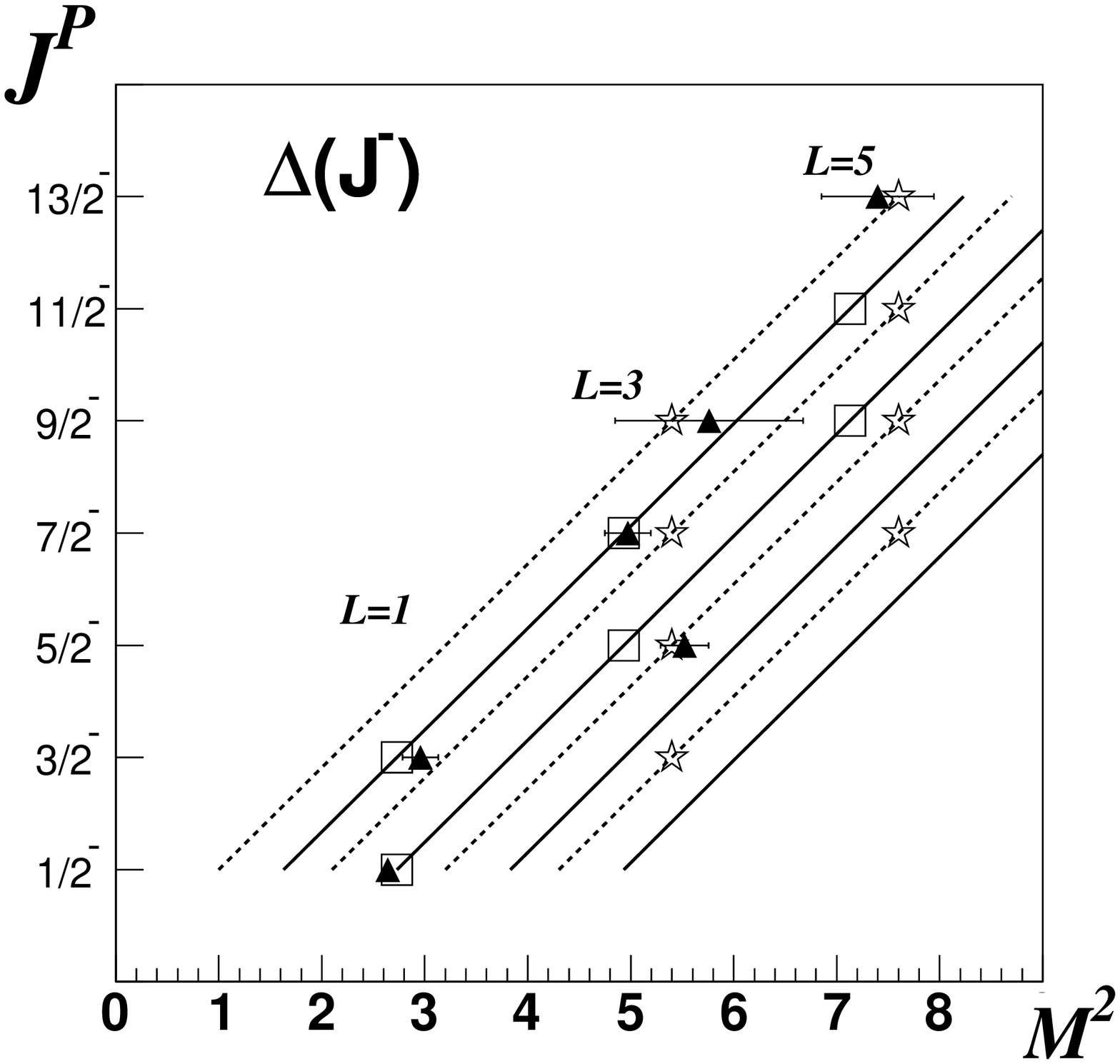,width=70mm}
             \epsfig{file=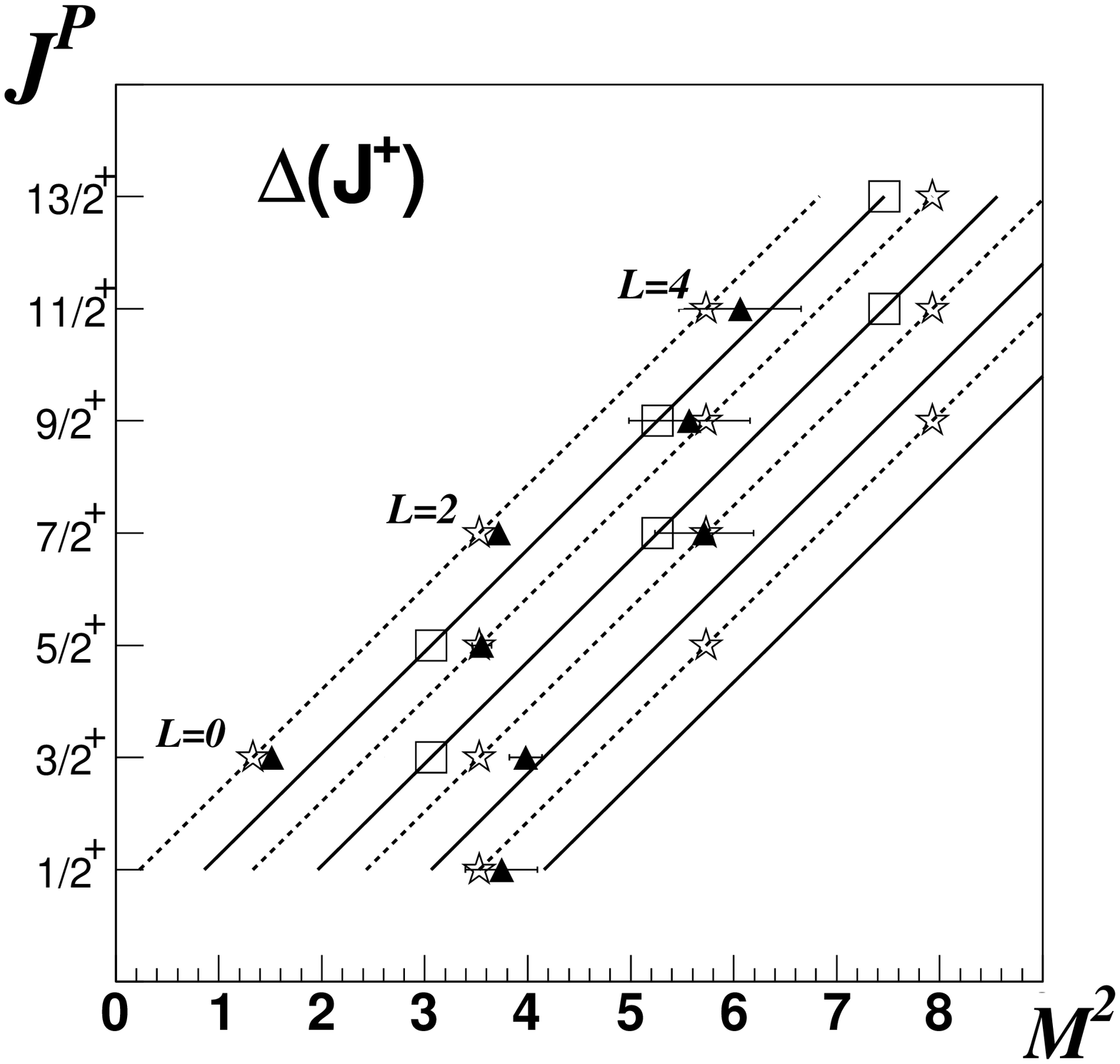,width=70mm}}
\centerline{\epsfig{file=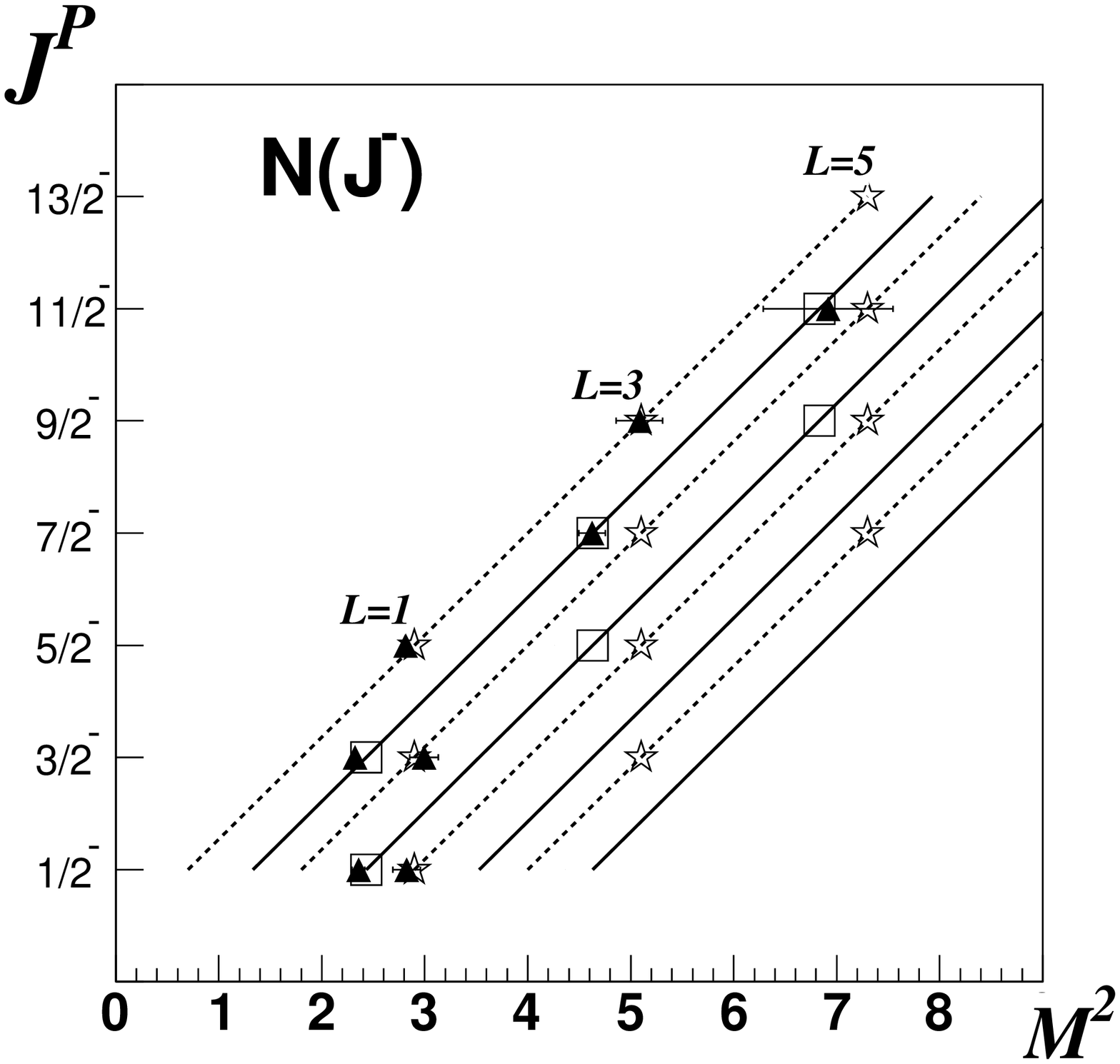,width=70mm}
             \epsfig{file=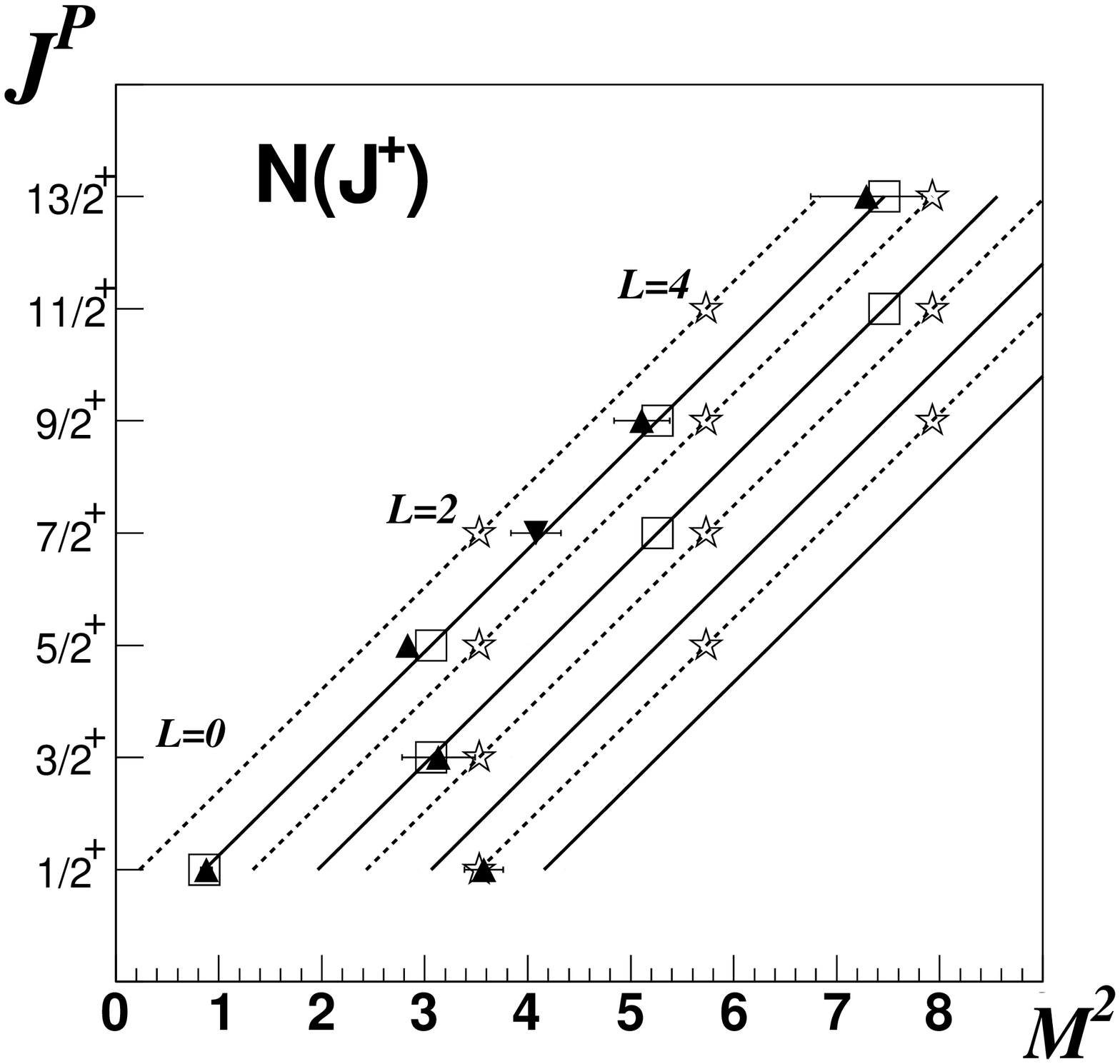,width=70mm}}
\caption{ Basic baryon settings   ($n=1$)  on
$(J^P,M^2)$ planes in the model with overlapping $qD^0_0(S=1/2)$ and
$qD^1_1(S=1/2)$ states (notations are as in Fig \ref{J-M-eq-M}.
}\label{J-M-eq-M1}
\end{figure}

In Fig. \ref{J-M-eq-M},
following  \cite{An,Al,Klempt,PDG}, we
show
$(J,M^2)$ plots as they look like
for  naive observers, while Fig.
\ref{J-M-eq-M1} demonstrates the $(J,M^2)$ plot for ground states
($n=1$) only.

We have three groups of states with $I=\frac 12$:
\bea
1)&& N_{J^P}\bigg(D^0_0;S=\frac 12,\,J=L-\frac 12\bigg),\quad
N_{J^P}\bigg(D^0_0;S=\frac 12,\,J=L-\frac 12\bigg),\\
2)&& N_{J^P}\bigg(D^1_1;S=\frac 12,\,J=L-\frac 12\bigg),\quad
N_{J^P}\bigg(D^1_1;S=\frac 12,\,J=L-\frac 12\bigg),\nn\\
3)&& N_{J^P}\bigg(D^1_1;S=\frac 32,\,J=L-\frac 32\bigg),\quad
N_{J^P}\bigg(D^1_1;S=\frac 32,\,J=L-\frac 12\bigg),\nn\\
&& N_{J^P}\bigg(D^1_1;S=\frac 32,\,J=L+\frac 12\bigg),\quad
N_{J^P}\bigg(D^1_1;S=\frac 32,\,J=L+\frac 32\bigg),\nn
\eea
and two with $I=\frac 32$:
\bea
1)&& \Delta_{J^P}\bigg(D^1_1;S=\frac 12,\,J=L-\frac 12\bigg),\quad
\Delta_{J^P}\bigg(D^1_1;S=\frac 12,\,J=L-\frac 12\bigg),\\
2)&& \Delta_{J^P}\bigg(D^1_1;S=\frac 32,\,J=L-\frac 32\bigg),\quad
\Delta_{J^P}\bigg(D^1_1;S=\frac 32,\,J=L-\frac 12\bigg),\nn\\
&& \Delta_{J^P}\bigg(D^1_1;S=\frac 32,\,J=L+\frac 12\bigg),\quad
\Delta_{J^P}\bigg(D^1_1;S=\frac 32,\,J=L+\frac 32\bigg),\nn
\eea
We suppose that in $I=\frac 12$ sector the states of groups (1) and (2)
with the same $J$ and $L$ have the equal masses -- resonances are
overlapping:
\bea
&& M_{N_{J^P}(D^0_0;S=\frac 12,\,J=L-\frac 12)}=
M_{N_{J^P}(D^1_1;S=\frac 12,\,J=L-\frac 12)}, \\
&& M_{N_{J^P}(D^0_0;S=\frac 12,\,J=L+\frac 12)}=
 M_{N_{J^P}(D^1_1;S=\frac 12,\,J=L+\frac 12}. \nn
\eea

\clearpage

\clearpage

\begin{figure}[h]
\centerline{\epsfig{file=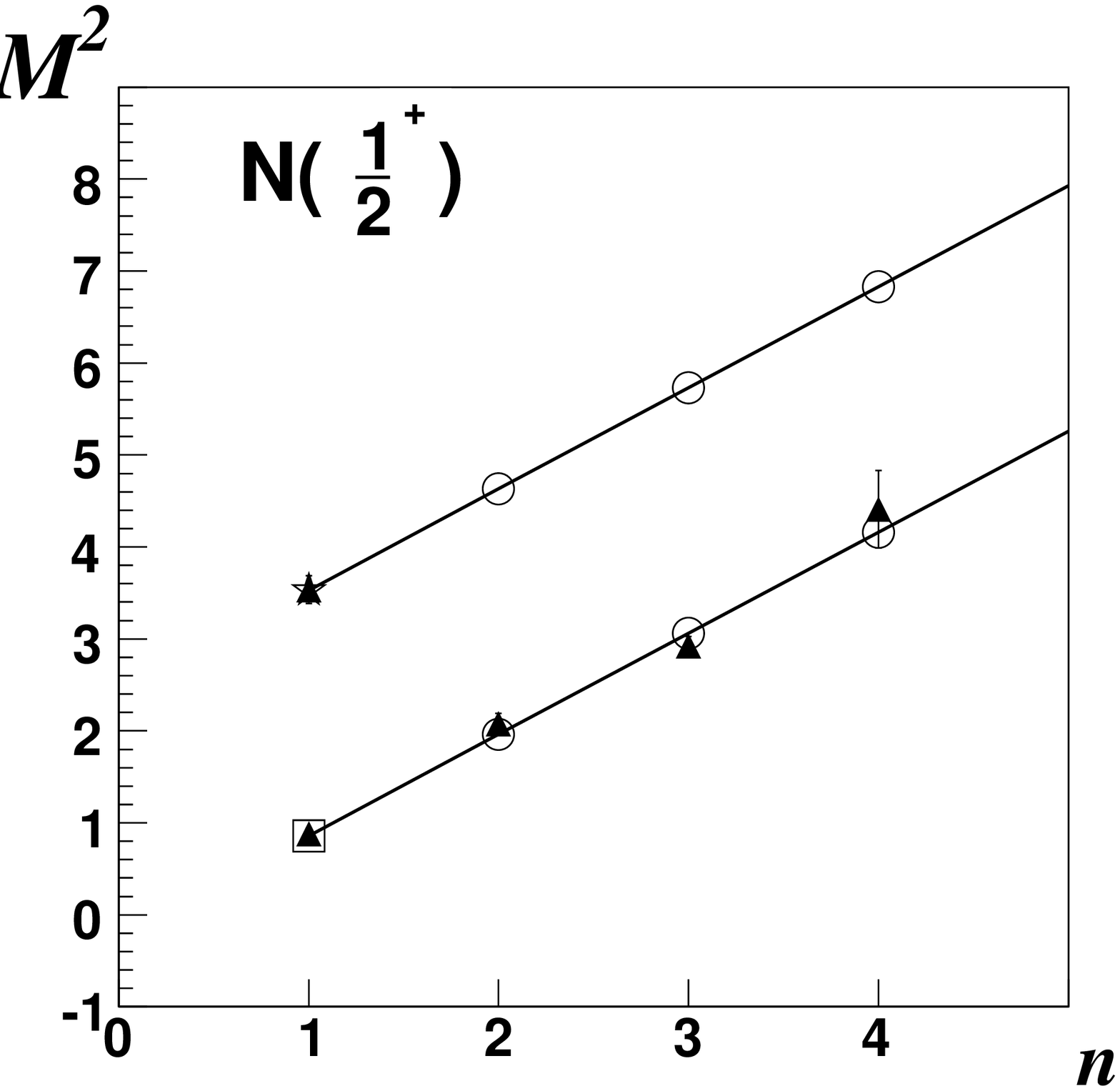,width=50mm}
             \epsfig{file=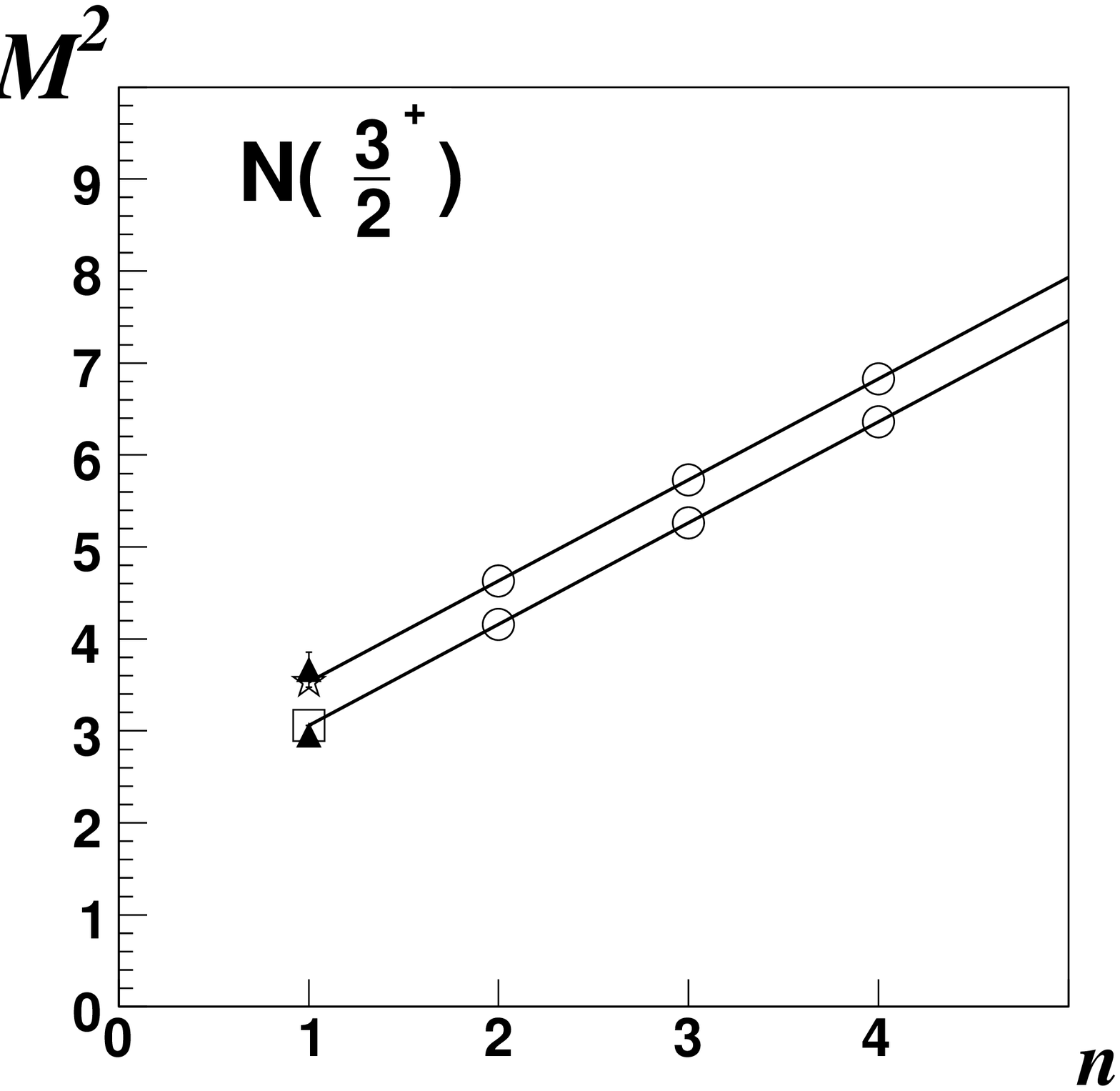,width=50mm}
             \epsfig{file=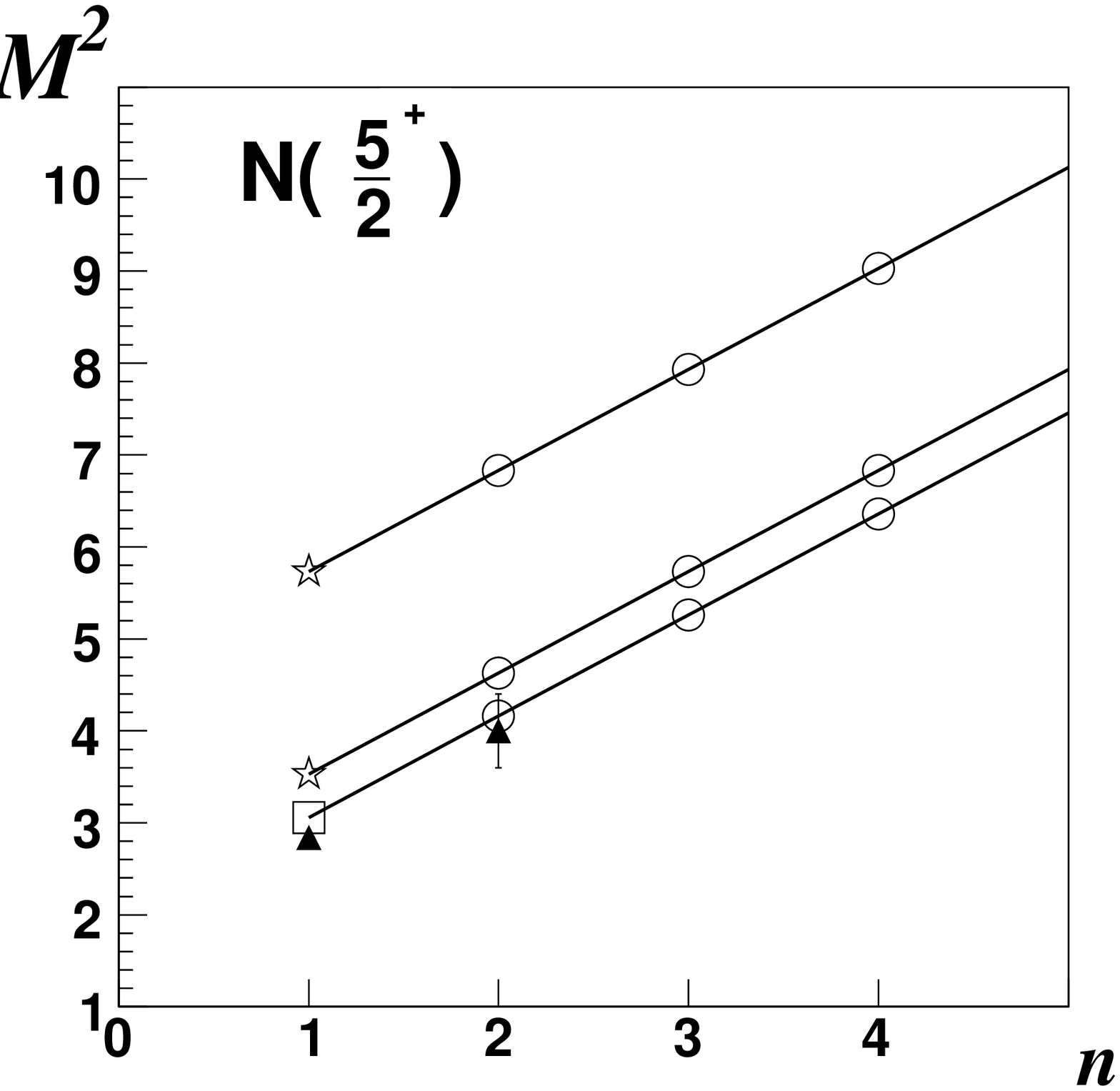,width=50mm}}
\centerline{\epsfig{file=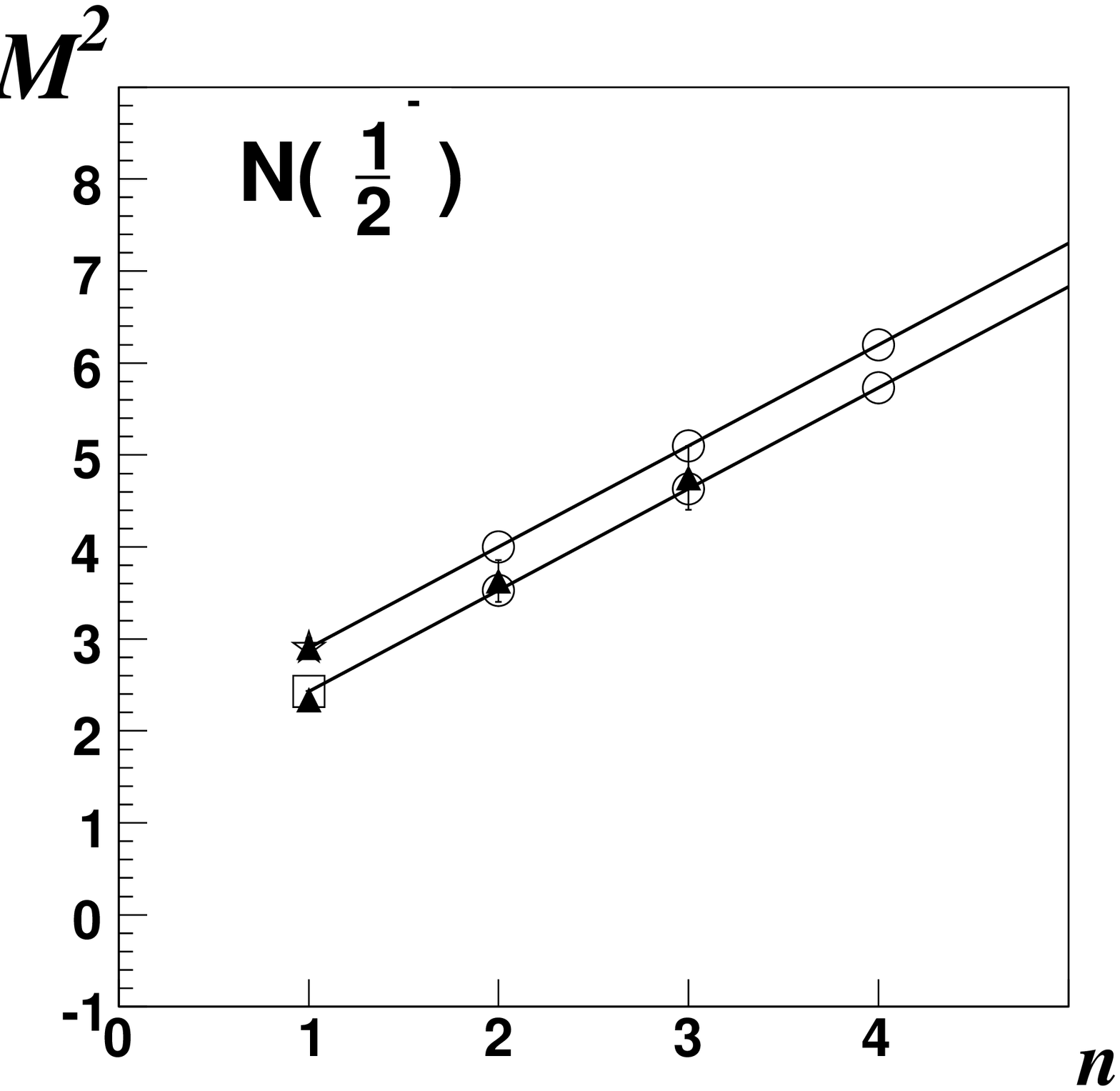,width=50mm}
             \epsfig{file=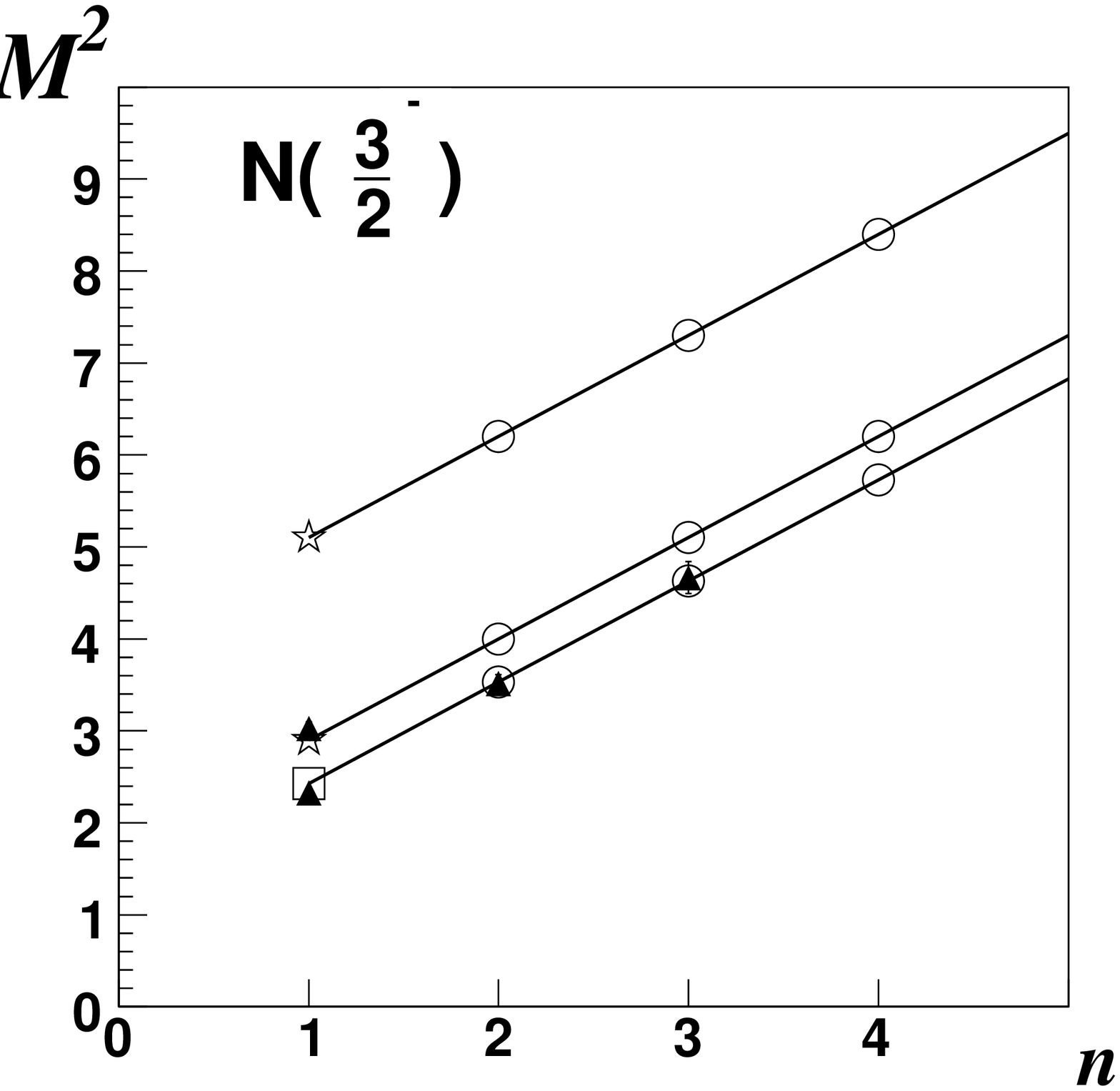,width=50mm}
             \epsfig{file=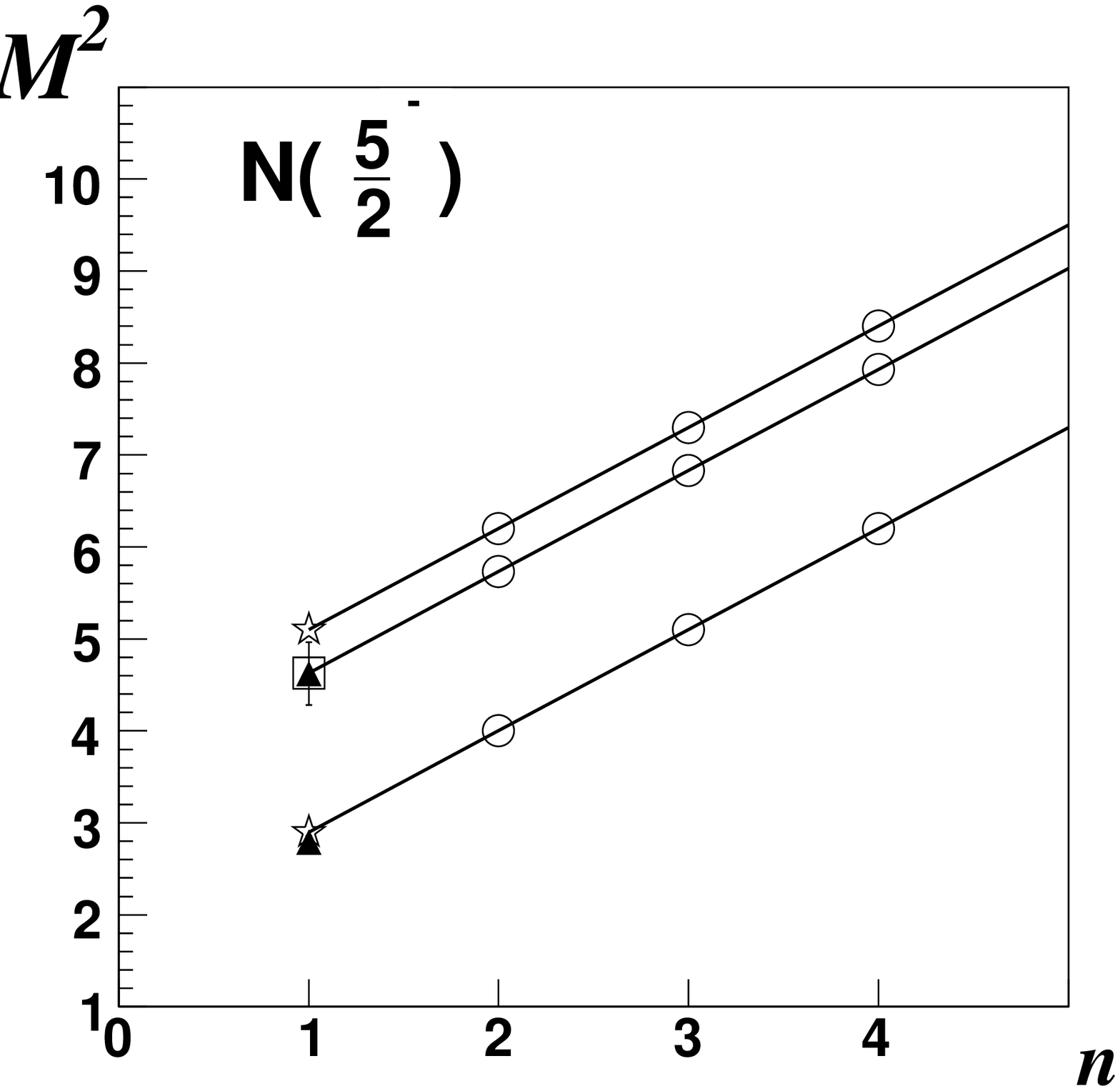,width=50mm}}
\centerline{\epsfig{file=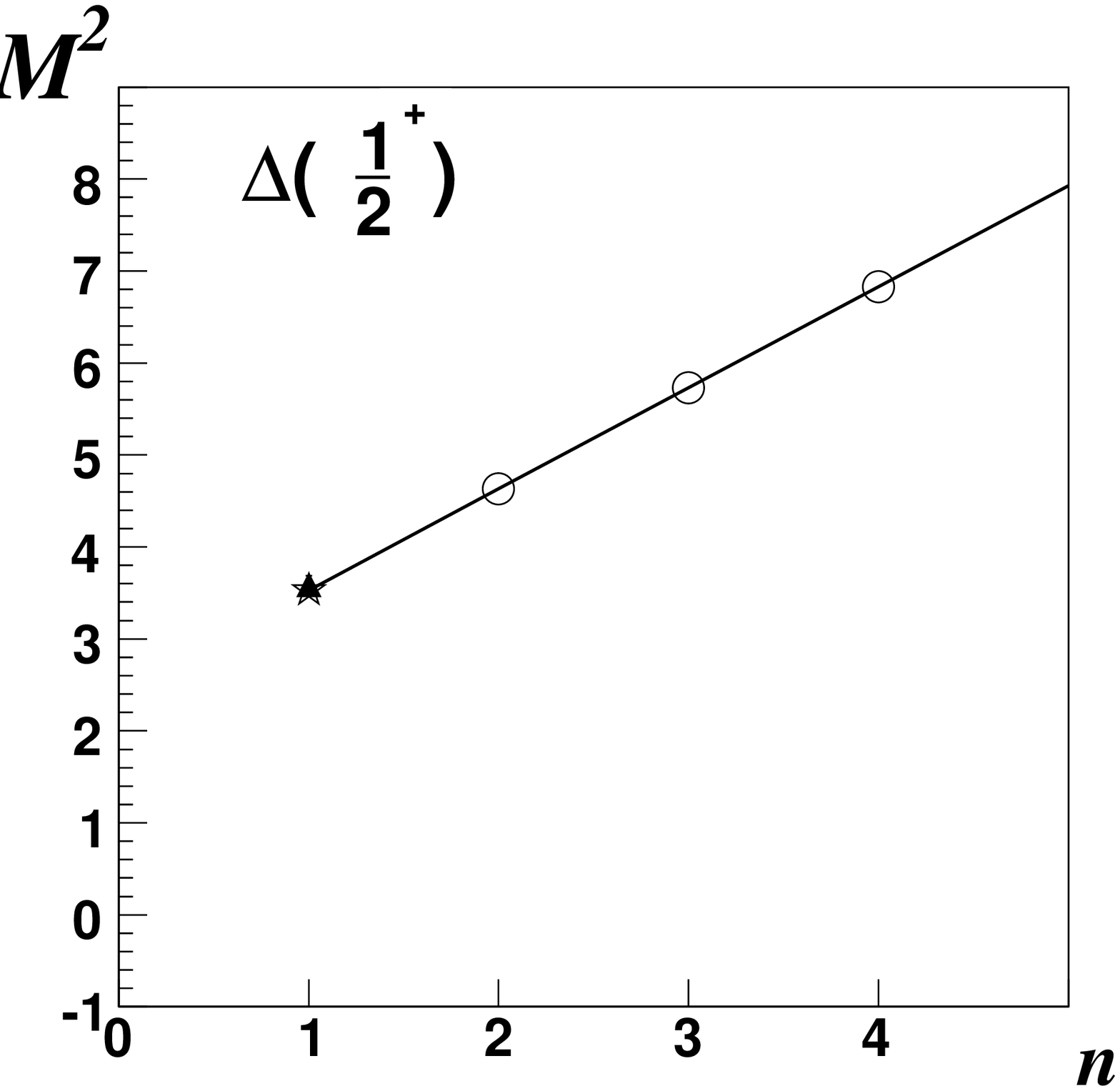,width=50mm}
             \epsfig{file=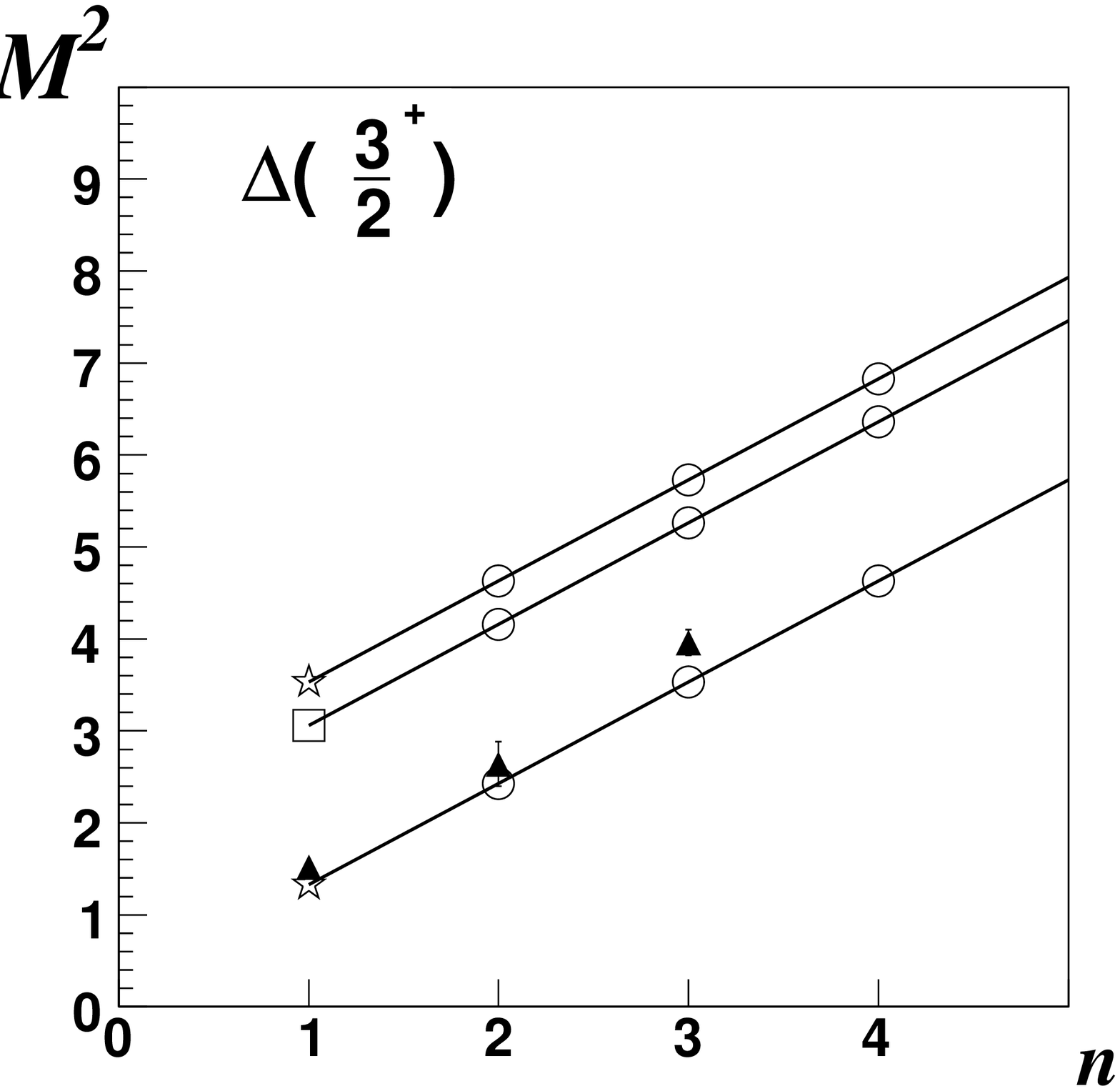,width=50mm}
             \epsfig{file=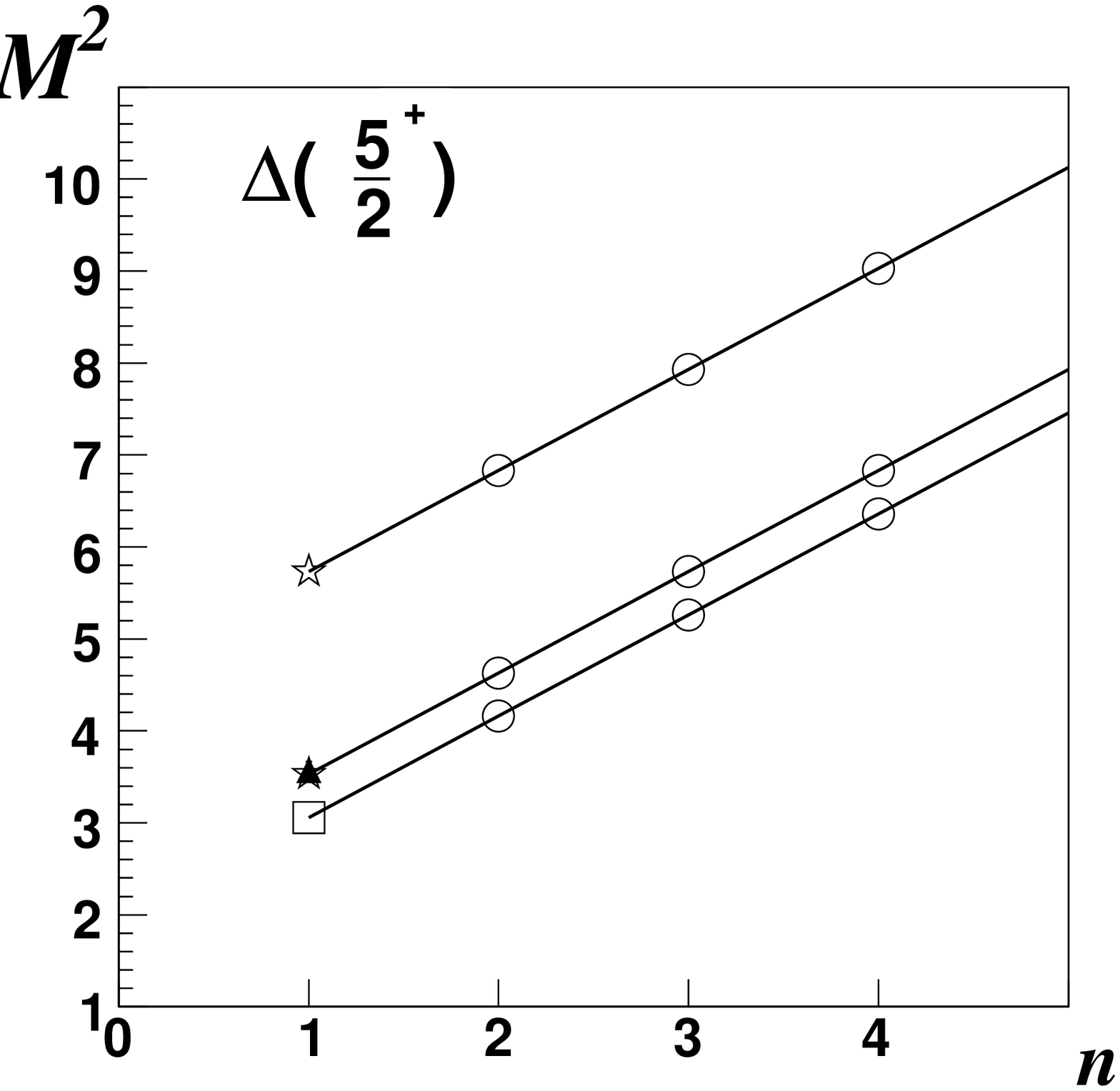,width=50mm}}
\centerline{\epsfig{file=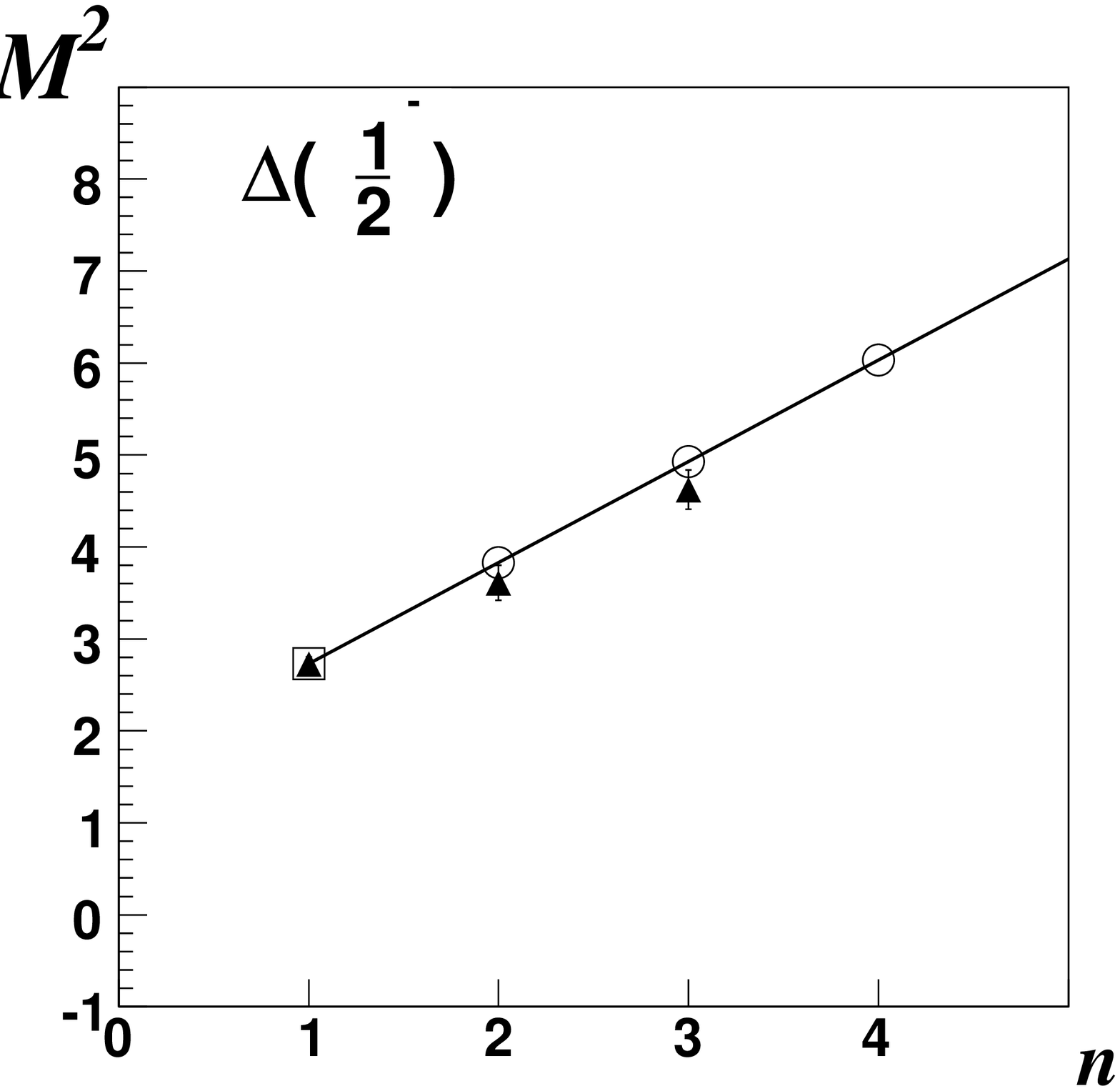,width=50mm}
             \epsfig{file=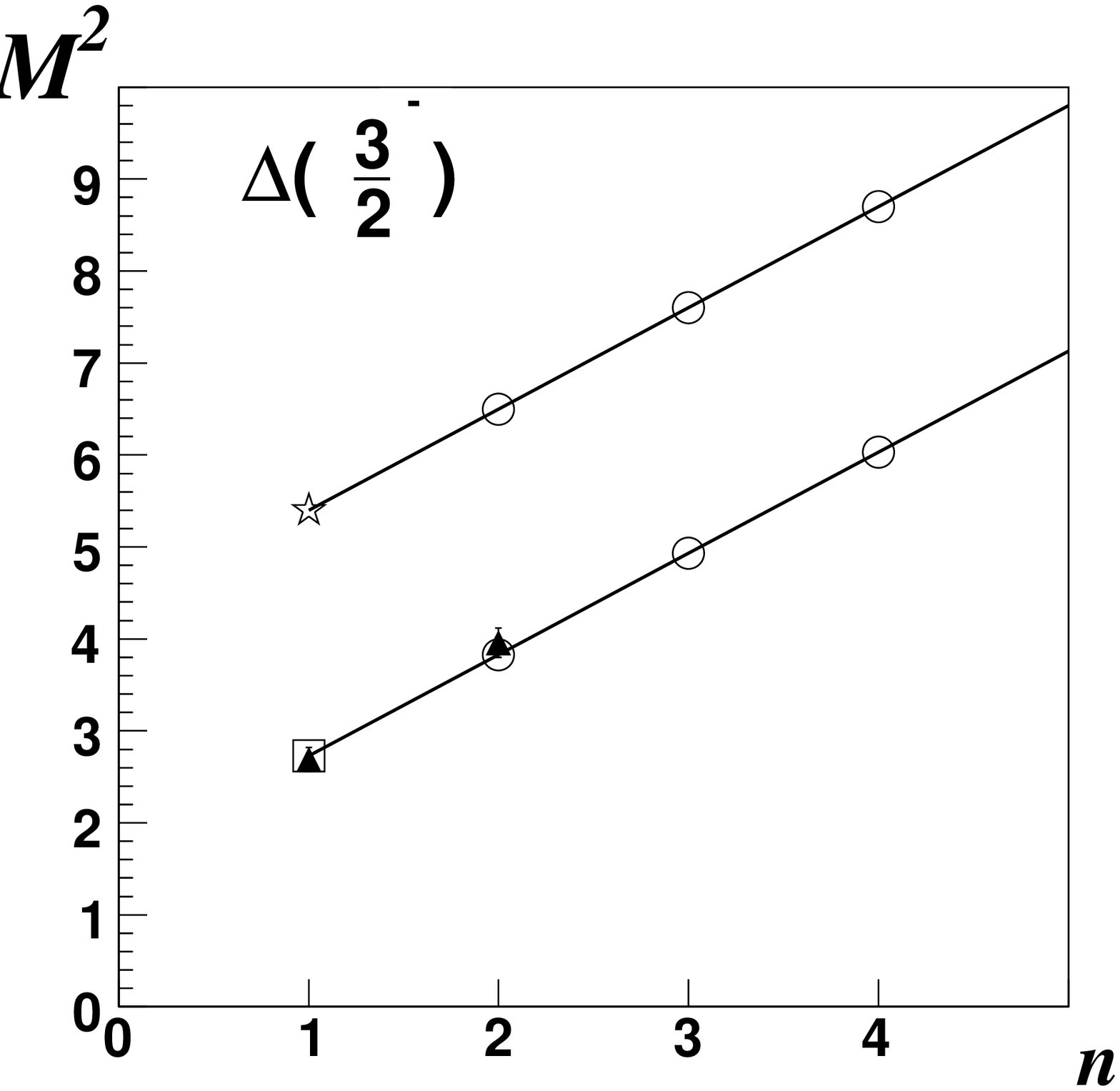,width=50mm}
             \epsfig{file=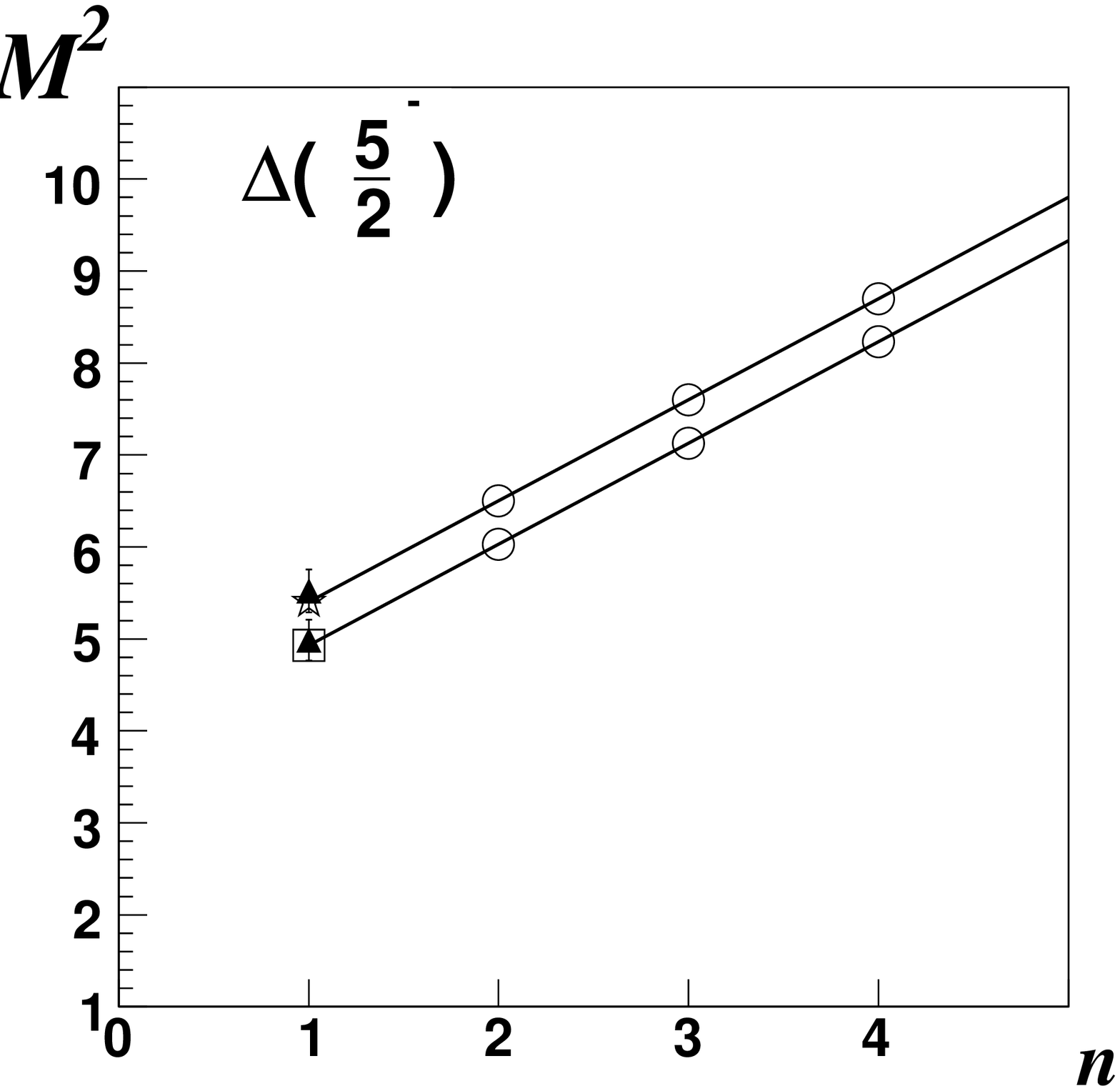,width=50mm}}
\caption{The $(n,M^2)$ plots for $I=1/2$ and $I=3/2$ states.}\label{n1}
\end{figure}

For $P=+$ states the values of baryon masses in the  $I=1/2$ sector are as follows:
\begin{equation}
{\renewcommand{\arraystretch}{0,5}
\begin{tabular}{l|l|l|llll}
$L=2$    &$S=\frac12$&    &   &$N(\frac32^+)$&$N(\frac52^+)$&      \\
     &     &$n=1$    &   &$(1770\pm 100)$&$(1685\pm  5)$&     \\
     &     &$n=2$    &   &$(1960\pm  30)$&$(2000\pm 100)$&     \\
     &     &$n=3$    &   &$\sim 2290$&$\sim 2290$&     \\
     &     &$n=4$    &   &$\sim 2520$&$\sim 2520$&     \\
    &$S=\frac32$&     &$N(\frac12^+)$&$N(\frac32^+)$&$N(\frac52^+)$&$N(\frac72^+)$   \\
     &     &$n=1$    &$(1890\pm 50)$&$\sim 1880$&$\sim 1880$&$(2020\pm 60)$  \\
     &     &$n=2$    &$\sim 2150$&$\sim 2150$&$\sim 2150$&$\sim 2150$  \\
     &     &$n=3$    &$\sim 2390$&$\sim 2390$&$\sim 2390$&$\sim 2390$  \\
     &     &$n=4$    &$\sim 2610$&$\sim 2610$&$\sim 2610$&$\sim 2610$  \\
\hline
$L=4$    &$S=\frac12$&     &   &$N(\frac72^+)$&$N(\frac92^+)$&      \\
     &     &$n=1$    &   &$\sim 2290$&$(2260\pm 60)$&     \\
     &     &$n=2$    &   &$\sim 2520$&$\sim 2520$&     \\
     &     &$n=3$    &   &$\sim 2730$&$\sim 2730$&     \\
       &$S=\frac32$&     &$N(\frac52^+)$&$N(\frac72^+)$&$N(\frac92^+)$&$N(\frac{11}2^+)$   \\
     &     &$n=1$    &$\sim 2390$&$\sim 2390$&$\sim 2390$&$\sim 2390$  \\
     &     &$n=2$    &$\sim 2610$&$\sim 2610$&$\sim 2610$&$\sim 2610$  \\
     &     &$n=3$    &$\sim 2820$&$\sim 2820$&$\sim 2820$&$\sim 2820$  \\
\hline
$L=6$    &$S=\frac12$&     &   &$N(\frac{11}2^+)$&$N(\frac{13}2^+)$&      \\
     &     &$n=1$    &   &$\sim 2730$&$(2700\pm 100)$&     \\
     &     &$n=2$    &   &$\sim 2930$&$\sim 2930$&     \\
       &$S=\frac32$&     &$N(\frac92^+)$&$N(\frac{11}2^+)$&$N(\frac{13}2^+)$&$N(\frac{15}2^+)$   \\
     &     &$n=1$    &$\sim 2820$&$\sim 2820$&$\sim 2820$&$\sim 2820$  \\
     &     &$n=2$    &$\sim 3000$&$\sim 3000$&$\sim 3000$&$\sim 3000$  \\
\end{tabular}
}
\end{equation}
and for negative-parity states:
\begin{equation}
{\renewcommand{\arraystretch}{0,5}
\begin{tabular}{l|l|l|llll}
$L=3$    &$S=\frac12$&     &   &$N(\frac52^-)$&$N(\frac72^-)$&      \\
     &     &$n=1$    &   &$(2160\pm 80)$&$(2150\pm 30)$&     \\
     &     &$n=2$    &   &$\sim 2390$&$\sim 2390$&     \\
     &     &$n=3$    &   &$\sim 2610$&$\sim 2610$&     \\
     &     &$n=4$    &   &$\sim 2820$&$\sim 2820$&     \\
\hline
    &$S=\frac32$&     &$N(\frac32^-)$&$N(\frac52^-)$&$N(\frac72^-)$&$N(\frac92^-)$   \\
     &     &$n=1$    &$\sim 2260$&$\sim 2260$&$\sim 2260$&$(2250\pm 50)$  \\
     &     &$n=2$    &$\sim 2490$&$\sim 2490$&$\sim 2490$&$\sim 2490$  \\
     &     &$n=3$    &$\sim 2700$&$\sim 2700$&$\sim 2700$&$\sim 2700$  \\
     &     &$n=4$    &$\sim 2900$&$\sim 2900$&$\sim 2900$&$\sim 2900$  \\
\hline
$L=5$    &$S=\frac12$&     &   &$N(\frac92^-)$&$N(\frac{11}2^-)$&      \\
     &     &$n=1$    &   &$\sim 2610$&$\sim 2610$&     \\
     &     &$n=2$    &   &$\sim 2820$&$\sim 2820$&     \\
     &     &$n=3$    &   &$\sim 3000$&$\sim 3000$&     \\
\hline
    &$S=\frac32$&     &$N(\frac72^-)$&$N(\frac92^-)$&$N(\frac{11}2^-)$&$N(\frac{13}2^-)$   \\
     &     &$n=1$    &$\sim 2700$&$\sim 2700$&$\sim 2701$&$\sim 2700$  \\
     &     &$n=2$    &$\sim 2900$&$\sim 2900$&$\sim 2898$&$\sim 2900$  \\
     &     &$n=3$    &$\sim 3080$&$\sim 3080$&$\sim 3082$&$\sim 3080$  \\
\hline
\end{tabular}
}
\end{equation}
In the $I=3/2$ sector we have for $P=+$ states:
\begin{equation}
{\renewcommand{\arraystretch}{0,5}
\begin{tabular}{l|l|l|llll}
$L=2$    &$S=\frac12$&     &   &$\Delta(\frac32^+)$&$\Delta(\frac52^+)$&      \\
     &     &$n=1$    &   &$\sim 1750$&$\sim 1750$&     \\
     &     &$n=2$    &   &$\sim 2040$&$\sim 2040$&     \\
     &     &$n=3$    &   &$\sim 2290$&$\sim 2290$&     \\
     &     &$n=4$    &   &$\sim 2520$&$\sim 2520$&     \\
\hline
    &$S=\frac32$&     &$\Delta(\frac12^+)$&$\Delta(\frac32^+)$&$\Delta(\frac52^+)$&$\Delta(\frac72^+)$   \\
     &     &$n=1$    &$(1935\pm 90)$&$(1995\pm 40)$&$(1885\pm 25)$&$(1928\pm 8)$  \\
     &     &$n=2$    &$\sim 2150$&$\sim 2151$&$\sim 2150$&$\sim 2150$  \\
     &     &$n=3$    &$\sim 2400$&$\sim 2400$&$\sim 2400$&$\sim 2400$  \\
     &     &$n=4$    &$\sim 2610$&$\sim 2610$&$\sim 2610$&$\sim 2610$  \\
\hline
$L=4$    &$S=\frac12$&     &   &$\Delta(\frac72^+)$&$\Delta(\frac92^+)$&      \\
     &     &$n=1$    &   &$\sim 2290$&$\sim 2290$&     \\
     &     &$n=2$    &   &$\sim 2520$&$\sim 2520$&     \\
     &     &$n=3$    &   &$\sim 2730$&$\sim 2730$&     \\
\hline
    &$S=\frac32$&     &$\Delta(\frac52^+)$&$\Delta(\frac72^+)$&$\Delta(\frac92^+)$&$\Delta(\frac{11}2^+)$   \\
     &     &$n=1$    &$\sim 2390$&$(2390\pm 100)$&$(2360\pm 125)$&$(2460\pm 120)$  \\
     &     &$n=2$    &$\sim 2610$&$\sim 2610$&$\sim 2610$&$\sim 2610$  \\
     &     &$n=3$    &$\sim 2820$&$\sim 2820$&$\sim 2820$&$\sim 2820$  \\
\hline
$L=6$    &$S=\frac12$&     &   &$\Delta(\frac{11}2^+)$&$\Delta(\frac{13}2^+)$&      \\
     &     &$n=1$    &   &$\sim 2730$&$\sim 2730$&     \\
     &     &$n=2$    &   &$\sim 2930$&$\sim 2930$&     \\
\hline
    &$S=\frac32$&     &$\Delta(\frac92^+)$&$\Delta(\frac{11}2^+)$&$\Delta(\frac{13}2^+)$&$\Delta(\frac{15}2^+)$   \\
     &     &$n=1$    &$\sim 2810$&$\sim 2810$&$\sim 2810$&$(2920\pm 100)$  \\
     &     &$n=2$    &$\sim 3000$&$\sim 3000$&$\sim 3000$&$\sim 3000$  \\
\hline
\end{tabular}
}
\end{equation}
and for $P=-$ ones:
\begin{equation}
{\renewcommand{\arraystretch}{0,5}
\begin{tabular}{l|l|l|llll}
$L=3$    &$S=\frac12$&     &   &$\Delta(\frac52^-)$&$\Delta(\frac72^-)$&      \\
     &     &$n=1$    &   &$\sim 2230$&$(2230\pm 50)$&     \\
     &     &$n=2$    &   &$\sim 2460$&$\sim 2460$&     \\
     &     &$n=3$    &   &$\sim 2670$&$\sim 2670$&     \\
     &     &$n=4$    &   &$\sim 2870$&$\sim 2870$&     \\
\hline
    &$S=\frac32$&     &$\Delta(\frac32^-)$&$\Delta(\frac52^-)$&$\Delta(\frac72^-)$&$\Delta(\frac92^-)$   \\
     &     &$n=1$    &$\sim 2320$&$(2350\pm 50)$&$\sim 2320$&$\sim 2320$  \\
     &     &$n=2$    &$\sim 2550$&$\sim 2550$&$\sim 2550$&$\sim 2550$  \\
     &     &$n=3$    &$\sim 2760$&$\sim 2760$&$\sim 2760$&$\sim 2760$  \\
     &     &$n=4$    &$\sim 2950$&$\sim 2950$&$\sim 2950$&$\sim 2950$  \\
\hline
$L=5$    &$S=\frac12$&     &   &$\Delta(\frac92^-)$&$\Delta(\frac{11}2^-)$&      \\
     &     &$n=1$    &   &$(2400\pm 190)$&$\sim 2670$&     \\
     &     &$n=2$    &   &$\sim 2870$&$\sim 2870$&     \\
     &     &$n=3$    &   &$\sim 3050$&$\sim 3050$&     \\
\hline
    &$S=\frac32$&     &$\Delta(\frac72^-)$&$\Delta(\frac92^-)$&$\Delta(\frac{11}2^-)$&$\Delta(\frac{13}2^-)$   \\
     &     &$n=1$    &$\sim 2760$&$\sim 2760$&$\sim 2760$&$(2720\pm 100)$  \\
     &     &$n=2$    &$\sim 2950$&$\sim 2950$&$\sim 2950$&$\sim 2950$  \\
     &     &$n=3$    &$\sim 3130$&$\sim 3130$&$\sim 3130$&$\sim 3130$  \\
\hline
\end{tabular}
}
\end{equation}
The above-written  equations allow us to predict the $(n,M^2)$ plots which
are shown in Fig. \ref{n1}.

\clearpage

\section{Conclusion}

Under the hypothesis of the quark--diquark structure of highly excited
baryons, we succeeded to suggest for them a realistic classification.
The introduction of diquarks gave a considerable reduction of excited
states; additional reduction was obtained owing to the assumption of the
overlapping of $(I=1/2)$ states with $S=1/2$.

Thus obtained a classification gave linear trajectories in the
$(J,M^2)$ and $(n,M^2)$ planes, which are strictly ordered. There is a
number of overlapping poles. The observation of two-pole and three-pole
structures in the complex-$M$ planes of partial amplitudes is a top-priority
task at the analysis of baryon spectra, while the next task consists in the
 writing and solving the spectral integral equation for quark--diquark
systems. Such an equation should be similar to that written and solved
before for the $q\bar q$ system \cite{si-qq}.

\section*{Acknowledgments}

We thank L.G. Dakhno
for helpful discussions. The paper was supported by the RFFI grants
07-02-01196-a and RSGSS-3628.2008.2.

 \end{document}